\documentclass[8pt,a4paper,usenatbib]{mnras}
\usepackage[utf8]{inputenc}
\usepackage{amsmath}
\usepackage{amsfonts}
\usepackage{amssymb}
\usepackage{graphicx}
\usepackage{lmodern}
\usepackage{multicol}
\usepackage{multirow}
\usepackage{hyperref}
\usepackage{comment}
\usepackage{verbatim}
\usepackage{pifont}
\usepackage{xspace}
\usepackage{xcolor}

\def\del#1{{}}

\hypersetup{
    colorlinks,
    linkcolor={red!50!red},
    citecolor={blue!50!blue},
    urlcolor={blue!80!black}
}

\def\s{{\rm s}} 
\def\yr{{\rm yr}} 

\def\age{{\rm age}}

\def\m{{\rm m}} 
\def\cm{{\rm c}\m} 
\def\km{{\rm k}\m} 
\def\pc{{\rm pc}} 
\def\kpc{{\rm k}\pc} 

\def\eV{{\rm eV}} 
\def\GeV{{\rm G}\eV} 
\def\TeV{{\rm T}\eV} 
\def\PeV{{\rm P}\eV} 
\def\erg{{\rm erg}} 

\def\deg{{\rm deg}} 

\def\G{{\rm G}} 
\def\muG{\mu\G} 
\def\kyr{{\rm kyr}} 

\def\de{{\rm d}} 

\def\free{{\rm free}}
\def\ST{{\rm ST}}
\def\ej{{\rm ej}}
\def\SN{{\rm SN}}
\def\swept{{\rm swept}}
\def\SNR{{\rm SNR}}

\newcommand{\ph}{\mathrm{ph}}

\def\de{\mathrm{d}}

\newcommand{\bs}[1]{\boldsymbol{#1}}
\newcommand{\AREPO}{\textsc{AREPO}\xspace}

\def\apjs{Astrophys. J. Suppl. Ser.}
\def\apjl{Astrophys. J. Lett.}
\def\aj{Astronom. J.}

\def\aap{Astron. Astrophys.}
\def\mnras{Mon. Not. R. Astron. Soc.}

\def\araa{ARA\&A}

\def\grl{Geophys. Res. Lett.}

\newcommand{\ls}{\mathrm{L}}

\newcommand{\vecbf}{\mathbfit}

\author[Pais et al.]
       {Matteo Pais,\thanks{E-mail: mpais@aip.de (MP), cpfrommer@aip.de (CP)}
         Christoph Pfrommer,
         Kristian Ehlert,
         Maria Werhahn,
         \newauthor
         Georg Winner\\
Leibniz-Institut f\"{u}r Astrophysik Potsdam,  An der Sternwarte 16, 14482 Potsdam, Germany \\
}

\title[Coherence scale of magnetic fields using TeV emission from SNRs]{Constraining the coherence scale of the interstellar magnetic field using TeV gamma-ray observations of supernova remnants} 
\date{}

\begin{document}
\maketitle


\begin{abstract}
Galactic cosmic rays (CRs) are believed to be accelerated at supernova remnant
(SNR) shocks. In the hadronic scenario the TeV gamma-ray emission from SNRs
originates from decaying pions that are produced in collisions of the
interstellar gas and CRs. Using CR-magnetohydrodynamic simulations, we show that
magnetic obliquity-dependent shock acceleration is able to reproduce the
observed TeV gamma-ray morphology of SNRs such as Vela Jr.\ and SN1006 solely by
varying the magnetic morphology. This implies that gamma-ray bright regions
result from quasi-parallel shocks (i.e., when the shock propagates at a narrow
angle to the upstream magnetic field), which are known to efficiently accelerate
CR protons, and that gamma-ray dark regions point to quasi-perpendicular shock
configurations. Comparison of the simulated gamma-ray morphology to observations
allows us to constrain the magnetic coherence scale $\lambda_B$ around Vela
Jr.\ and SN1006 to $\lambda_B \simeq 13_{-4.3}^{+13}$~pc and $\lambda_B
>200_{-10}^{+80}$~pc, respectively, where the ambient magnetic field of SN1006
is consistent with being largely homogeneous. We find consistent pure
  hadronic and mixed hadronic-leptonic models that both reproduce the
  multi-frequency spectra from the radio to TeV gamma rays and match the
  observed gamma-ray morphology. Finally, to capture the propagation of a SNR
  shock in a clumpy interstellar medium, we study the interaction of a shock
  with a dense cloud with numerical simulations and analytics. We construct an
  analytical gamma-ray model for a core collapse SNR propagating through a
  structured interstellar medium, and show that the gamma-ray luminosity is only
  biased by 30\% for realistic parameters.
\end{abstract}

\begin{keywords}
Magnetohydrodynamics (MHD) -- ISM: supernova remnants -- cosmic rays -- gamma-rays 
\end{keywords}

\section{Introduction}
SNR shocks energize Galactic CRs via diffusive shock acceleration.  This
converts about $5-10\%$ of the kinetic energy into a non-thermal, power-law
momentum distribution of CRs \citep{1978MNRAS.182..147B,1978ApJ...221L..29B}.
The most direct observational evidence for this is the GeV and TeV gamma-ray
emission from SNRs. There are two competing models: in the leptonic model, CR
electrons Compton upscatter (interstellar) radiation fields whereas in the
hadronic model inelastic collisions between the interstellar medium (ISM) and
CRs produce neutral pions that decay into gamma rays \citep{2009ARA&A..47..523H,
  2010ApJ...708..965Z}. The latter process produces a kinematic spectral feature
below GeV energies, as recently observed by the Fermi gamma-ray telescope
\citep{2013Sci...339..807A}. However, the steep high-energy spectral slope
raises questions whether this represents an unambiguous proof of CR hadron
acceleration at this SNR \citep{2016A&A...595A..58C}.
\vspace{1em}

The hadronic model requires efficient CR hadrons acceleration, which must be
accompanied by substantial magnetic field amplification via the hybrid
non-resonant instability \citep{2004MNRAS.353..550B}. This finding received
strong observational support with the detection of thin X-ray synchrotron
filaments at several SNR shocks. Those filaments exhibit fast (year-scale)
variability and likely result from cooling of freshly accelerated electrons in
magnetic fields of $\approx1\,$mG \citep{2007Natur.449..576U}. Spatial
correlations between gamma-ray brightness and gas column density are another
consequence of the hadronic model and are expected for core-collapse supernovae,
which explode inside molecular clouds due to the fast evolution of their massive
progenitor stars. Combining synchrotron and inverse Compton emission in the
leptonic model yields volume-filling magnetic field strengths of
$\approx10\,\mu$G \citep{2016EPJWC.12104001G}, which are compatible with
mG-field strengths inferred from X-ray synchrotron filaments only when assuming
a clumpy medium. It has also been argued that a rising gamma-ray energy spectrum
with increasing photon energy provides evidence for leptonic models. However,
such a spectrum can also be obtained in the hadronic model when considering a
clumpy ISM because of proton propagation effects that substantially harden the
proton spectrum inside dense clumps in comparison to the acceleration spectrum
in the diffuse ISM \citep{2014MNRAS.445L..70G, 2019MNRAS.487.3199C}.  The
absence of thermal X–rays from the remnant provides additional support for this
scenario because the shock will considerably slow down while penetrating into
dense clumps and thus cannot heat them up to X-ray emitting temperatures
\citep{2012ApJ...744...71I}.

Alternatively, leptonic scenarios have been proposed to explain the gamma-ray
emission from young SNRs in an ISM model with homogeneous density
\citep[e.g.,][]{1996A&A...307L..57P} in order to overcome the lack of thermal
X-ray emission. This results in a low upper limit on the ISM density. Leptonic
models evoke inverse Compton (IC) scattering of CR electrons with a photon field
that is provided by the ubiquitous cosmic microwave background in combination
with (dust-processed) stellar light.  For example, \citet{2016ApJ...823...44X}
uses leptonic models to explain the emission from the South-Western limb of
SN1006 while this explanation is extended to the entire remnant
\citep{2011MNRAS.413.1643P, 2012MNRAS.425.2810A}.  The leptonic scenario
naturally explains the correlation between X-ray synchrotron and IC gamma-ray
emitting regions such as in Vela Jr.\ \citep{2007ApJ...661..236A} and matches
the observed broadband spectrum. On the other hand, this model implies low
magnetic field strengths, which are in contradiction to the narrow filamentary
structures detected in X-rays. Hence, we need to understand the detailed spatial
structure of SNRs across different wave lengths to unambiguously identify
emission and particle acceleration processes.

The high angular resolution ($<0.1^\circ$) of imaging air \v{C}erenkov
telescopes H.E.S.S., VERITAS, and MAGIC enables detailed morphological gamma-ray
studies of SNRs and to separate or exclude contributions by compact sources such
as pulsars. In particular, TeV gamma-ray observations have delivered a rich
morphology of shell-type SNRs, ranging from the bi-lobed emission of SN1006
\citep{2010AA...516A..62A} to the filamentous, patchy appearances of Vela
Jr.\ \citep{2018AA...612A...7H} and RX-J1713 \citep{2018A&A...612A...6H}, to
  the young, type Ia SNR Tycho G120.1+01.4 \citep{2017ApJ...836...23A}. In
  principle, the patchy gamma-ray morphology could result from density
inhomogeneities \citep{2008A&A...492..695B,2000A&A...355..211A} of the ambient
ISM. It yet remains to be seen whether the fluctuation amplitude necessary for
the observed gamma-ray patchiness does not introduce a corrugated shock surface
\citep{2016MNRAS.463.3989J} that is inconsistent with the observed spherical
blast wave.

Here, we propose a different model in which the acceleration process imprints a
rich gamma-ray morphology due to the global magnetic morphology
\citep{2018MNRAS.478.5278P}. Hybrid particle-in-cell simulations of
non-relativistic, strong shocks show that diffusive shock acceleration of hadrons
efficiently operates for quasi-parallel configurations (i.e., when the shock
propagates along the upstream magnetic field or moves at a narrow angle to it)
and converts around 15\% of the available energy to CRs
\citep{2014ApJ...783...91C}. In contrast, a shock that propagates perpendicular
to the magnetic field (or at a large angle to it, i.e., a quasi-perpendicular
configuration) is an inefficient accelerator without strong pre-existing
  turbulence at the CR gyroscale \citep{1992GeoRL..19..433G} because charged
particles are bound to gyrate around the flux-frozen magnetic field. As the
magnetized plasma sweeps past the shock, so are the gyrating particles, which
cannot return back upstream.

We aim to explain the apparently disparate TeV morphologies of SNR 1006 and Vela
Jr.\ within a single physical model assuming a hadronic model linking gamma-ray
morphology and local orientation of the magnetic field. To this end, we run a
suite of simulations modeling a point explosion that encounters a range in
magnetic field morphologies, from a homogeneous field to a mixture of
homogeneous and turbulent fields to fully turbulent fields with varying
coherence lengths in Section~\ref{sec:sims}. We rescale our simulation
parameters within observational limits to reproduce the observed gamma-ray
spectra and flux (Section~\ref{sec: Multifreq}). Comparing simulated to observed
morphologies in Section~\ref{sec:morphological_modelling} allows to constrain
the magnetic coherence length that the unperturbed ISM had before it encountered
the SNR blast wave. Assuming statistical homogeneity, we thus constrain the
magnetic coherence scale in the immediate vicinity of the SNR. We explore how
the gamma-ray signal is modified by launching the SNR into stellar wind profiles
instead of a homogeneous ISM environment in Section~\ref{sec: stellar wind} and
conclude in Section~\ref{sec:conclusion}. In Appendix~\ref{sec: appx}, we
  study the interaction of a shock with a dense cloud with numerical simuations
  and analytics and construct an analytical model for the gamma-ray luminosity
  from a core collapse SNR that interacts with a structured ISM with a large
  population of dense, cold clouds.

\section{Simulation setup}
\label{sec:sims}

\subsection{Rationale}

We aim to produce realistically looking TeV gamma-ray maps with the least
  amount of necessary physical complexity but including all the required
  processes for drawing transparent and robust conclusions from our
  three-dimensional MHD simulations. To this end we follow the following
  rationale:
\begin{itemize}
\item The general numerical setup consists of simulating the Sedov-Taylor phase
  of SNRs by injecting energy at the initial time into our simulation domain
  that is filled with magnetized plasma with different correlation lengths. We
  identify the shock during the run-time of the simulations, inject CR energy at
  the shock with an efficiency that is in agreement with the results of ab
  initio particle-in-cell plasma simulations, and advect the CR energy density
  with the thermal plasma.
\item In order to explain the TeV gamma-ray morphology for two famous shell-type
  SNR, SN1006 and Vela Jr., we have to realistically model the external medium
  that the remnant shocks are propagating into. Because the thermonuclear
  supernova SN1006 (of type Ia) is located at a Galactic height of 0.4~kpc, the
  shock propagates into low-density medium and encounters mostly homogeneous
  magnetic field that was potentially stretched due to a galactic outflow or the
  Parker instability.
\item In contrast, the Vela Jr.\ SNR is thought to be associated with a
  core-collapse supernova explosion \citep{2002ApJ...574..155W}, which results
  from the collapse of a massive star. Hence, in the post-processing, we account
  for the free expansion phase preceeding the Sedov-Taylor phase and rescale the
  final radii accordingly. The Vela Jr.\ SNR encounters a highly structured,
  multiphase ISM that is typical of star forming regions. This is modelled with
  a population of dense gaseous clumps embedded in a nearly homogeneous
  background medium, which simultaneously explains the absence of thermal X-ray
  emission and hard TeV gamma-ray spectra in the hadronic model
  \citep{2019MNRAS.487.3199C}. Finally, we additionally simulate the expansion
  of the SNR into two different stellar wind profiles that bracket the
  uncertainty in the progenitor model and study its impact on the TeV gamma-ray
  morphology.
\end{itemize}

\subsection{Simulation setup}

We perform our simulations with the second-order accurate, adaptive moving-mesh
code \AREPO \citep{2010MNRAS.401..791S,2016MNRAS.455.1134P}, using standard
parameters for mesh regularization. Magnetic fields are treated with ideal
magnetohydrodynamics \citep{2013MNRAS.432..176P}, using the Powell scheme for
divergence control \citep{1999JCoPh.154..284P}. CRs are modelled as a
relativistic fluid with adiabatic index $4/3$ in a two-fluid approximation
\citep{2017MNRAS.465.4500P}.

We localize and characterize shocks during the simulation
\citep{2015MNRAS.446.3992S} to inject CRs into the downstream
\citep{2017MNRAS.465.4500P} with an efficiency that depends on the upstream
magnetic obliquity \citep{2018MNRAS.478.5278P}. We adopt a maximum acceleration
efficiency for CRs at quasi-parallel shocks of $15\%$ that approaches zero for
quasi-perpendicular shocks \citep{2014ApJ...783...91C}. We only model the
dominant advective CR transport and neglect CR diffusion and streaming. This is
justified since diffusively shock accelerated CRs experience efficient Bohm
diffusion with a coefficient $\kappa_\rmn{Bohm}$ at the shock
\citep{2006NatPh...2..614S}, which implies a CR precursor ($L_\rmn{prec}$) that
is smaller than our grid cells,
$L_\rmn{prec}\sim\sqrt{\kappa_\rmn{Bohm}t}\sim0.1~\rmn{pc}\times(pc/10\,\rmn{TeV})^{1/2}\,
(B/100\,\mu\rmn{G})^{-1/2}\,(t/10^3\,\rmn{yr})^{1/2}$. We neglect slow
non-adiabatic CR cooling processes in comparison to the fast Sedov expansion.

Each simulation follows a point explosion that results from depositing
$E_{\rmn{SN}}=10^{51}~\erg$ in a homogeneous periodic box
\citep{2017MNRAS.465.4500P}. This leads to an energy-driven,
spherically-symmetric strong shock expanding in a low-pressure ISM with mean
molecular weight $\mu=1.4$. Our initial conditions are constructed by first
generating a Voronoi mesh with randomly distributed mesh-generating points in
our three-dimensional simulation box with $200^3$ cells that we then relax via
Lloyd's algorithm \citep{Lloyd82} to obtain a glass-like configuration.

To simulate a realistic star forming environment for Vela Jr. we inserted
  $7\times 10^3$ uniformly distributed small, dense clumps with a number
  density of $n_\rmn{c} = 10^3~\cm^{-3}$ and a diameter of $0.1~\pc$
  \citep{1977ApJ...218..148M}. More details about the clumps can be found in
  Appendix~\ref{sec: appx}.

Our turbulent magnetic fields exhibit magnetic power spectra of Kolmogorov type
with different coherence lengths. The three magnetic field components are
treated independently so that the resulting field has a random phase. To fulfill
the constraint $\bs{\nabla}\bs{\cdot}\vecbf{B}=0$ we project out the radial field
component in Fourier space. We assume a low ISM pressure of 0.44~\eV~\cm$^{-3}$
and scale the field strength to an average plasma beta factor of unity. To
ensure pressure equilibrium in the initial conditions, we adopt temperature
fluctuations of the form $nk_\rmn{B}\delta{}T=-\delta\vecbf{B}^2/(8\pi)$ (for 
details, see \citealt{2018MNRAS.478.5278P}).

In our physical set-up, there are two different processes driving
turbulence. The process of diffusive shock acceleration excites non-linear
(turbulent) Bell modes on scales below the CRs' gyroradii
\citep{2004MNRAS.353..550B}. Because we adopt the obliquity-dependent CR
acceleration efficiency from self-consistent plasma simulations
\citep{2014ApJ...783...91C} in our sub-grid model, we implicitly account for the
full kinetic physics. 

On the contrary, the magnetic turbulence that we
explicitly model in our simulations reflects the supernovae-driven ISM
turbulence with varying injection scales from 4 to 200~pc. The magnetic
fluctuations cascade down to levels of $\delta B / B \approx 10^{-3}$ at
resonant length scales of TeV CRs so that they do not interfere with the
large-scale field topology at the shock. To derive this result, we assumed
Alfv\'enic turbulence for parallelly propagating Alfv\'en waves according to
theory of magnetohydrodynamical turbulence \cite{1995ApJ...438..763G} in a mean
magnetic field of $10\,\mu$G. Hence, the small fluctuation amplitude and the
enormous scale separation of injection-to-gyroscale of $\approx10^5$ justifies
our separate treatment of these two processes.

\subsection{Observational modeling}
\label{sec:obs}

In order to connect our simulations to gamma-ray observations, we need to take
into account all observational constraints on ISM properties. For practical
reasons, here we derive approximate scaling laws of the gamma-ray flux in the
Sedov-Taylor regime and adopt the simplified assumption of a CR spectrum with
index $\alpha_\rmn{p}=2$, which yields an equal contribution to the total CR
energy for each decade in CR momentum. These scaling relations enable us to find
parameter combinations of the SNR and surrounding ISM, which match observed
gamma-ray fluxes in the hadronic model. We will vary those parameter
combinations in a detailed multi-frequency analysis in Section~\ref{sec:
  Multifreq} study how they vary if the SNR shock propagates in a stellar wind
profile in Section~\ref{sec: stellar wind}.

\subsubsection{SN1006}
SN1006 represents a good case study of a SNR with a unique gamma-ray
morphology. Moreover, its exactly known age allows to tightly constrain other
environmental ISM parameters. Low-resolution HI measurements of the density
around SN1006 suggest a diffuse density of $n = 0.3~\cm^{-3}$ and an interaction
of the SNR with a dense cloud of $n = 0.5~\cm^{-3}$
\citep{2002A&A...387.1047D}. Studies based on X-ray spectroscopy estimate the
density of the North-Western rim of the SNR to be $n = (0.15 - 0.25)~\cm^{-3}$
\citep{2003ApJ...586.1162L} which is consistent with hydrodynamic simulations of
an explosion energy of $10^{51}~\erg$ in a homogeneous medium
\citep{2001ApJ...549.1119W}. More recent papers suggest an even lower ambient
density for the North-Western rim of $n=0.085~\cm^{-3}$, which is estimated
based on X-ray proper motion measurements \citep{2009ApJ...692L.105K}, down to a
density of $n=0.05~\cm^{-3}$ for the South-Eastern rim, which is based on X-ray
observations in combination with a shock-plasma model
\citep{2007AA...475..883A}.  Such low densities, however, would require an
uncomfortably high explosion energy in order to explain the gamma-ray emission
of SN1006 in the hadronic model. Assuming a CR proton acceleration efficiency of
$10\%$ yields $E_{\rmn{SN}}=3\times 10^{51}~\erg$ \citep{2010AA...516A..62A}.

Integrating the differential gamma-ray flux \citep[equation (1)
  of][]{2016EPJWC.12104001G} yields an estimate of the TeV gamma-ray flux,
$\mathcal{F}_\gamma$, of a SNR in the hadronic model:
\begin{equation}
    \mathcal{F}_\gamma \simeq  \int_{1~\TeV}^{100~\TeV} F_\gamma \de E_\gamma=
    \int_{1~\TeV}^{100~\TeV} \dfrac{ 4 W_\mathrm{p} ~E_\gamma^{-2} \de E_\gamma}{\ln(E_\mathrm{max}/ E_\mathrm{min}) \tau_{\pi^0} 4\pi D^2}
\end{equation}
where $ \tau_{\pi^0} \simeq 1.6 \times 10^{9} \left( n / 0.1~\cm^{-3}
\right)^{-1}~\yr$ is the energy loss time due to neutral pion production, $n$ is
the ISM number density assuming cosmic abundances ($\mu = 1.4$), 
$D$ is the distance to the SNR, $W_\mathrm{p}$ is the total
proton energy and $E_\gamma$ is the gamma-ray energy. Here, we assume that the
CR spectrum extends from $E_\mathrm{min} = 1~\GeV$ to $E_\mathrm{max} = 4~\PeV$,
which corresponds to the energy of the knee. The factor of $4$ accounts for the
compression of the density at the shock. The integration yields
\begin{equation}
  \label{eq: flux}
  \mathcal{F}_\gamma \simeq  2.7 \times 10^{-12} \left( \dfrac{W_\mathrm{p}}{10^{50}\,\erg}\right)
  \left( \dfrac{n}{0.1\,\cm^{-3}}\right) \left( \dfrac{D}{1\,\kpc} \right)^{-2}\!\! \dfrac{\ph}{\cm^{2}~\s}.
\end{equation}
The self-similar solution for a strong shock in the Sedov-Taylor regime
\citep{1959sdmm.book.....S} states that the shock radius $r_\ST$ evolves as
  \begin{equation}
  r_\ST(t) = \left(\dfrac{E_\mathrm{SN}}{\alpha \rho}\right)^{1/5} t^{2/5}_\age,
  \end{equation}
  where $\rho$ is the ISM mass density, $t_\age$ is the age of the remnant and
    $\alpha$ a dimensionless factor depending on the adiabatic index of the
    fluid. For a mixture of thermal gas and freshly accelerated CRs with a
    maximum efficiency of $15\%$ we find $\alpha=0.52$
    \citep{2018MNRAS.478.5278P}.  The resulting shock radius for typical ISM
    parameters is
  \begin{equation}
  \label{eq: sedov}
  r_\ST = 7.82~\pc \left( \dfrac{E_\mathrm{SN}}{10^{51}~\erg}\right)^{1/5} \left( \dfrac{n }{0.1~\cm^{-3}}\right)^{-1/5}
  \left( \dfrac{t_\age}{1000~\yr}\right)^{2/5}.
  \end{equation}
  The shock radius can be expressed by the angle it subtends on the sky
    (assuming the small-angle approximation)
\begin{equation}
\label{eq: angular}
r_\ST = D \sin\left(\dfrac{\theta}{2}\right) \simeq 8.7~\pc \left( \dfrac{D}{1~\kpc}\right) \left(\dfrac{\theta}{\deg} \right).
\end{equation}
  Combining Eq.~\eqref{eq: sedov} with Eq.~\eqref{eq: angular} we derive the
    following formula for the density of SN1006:
  \begin{equation}
  \label{eq: density}
n = 0.1~\cm^{-3} \left( \dfrac{t_\age}{1~\kyr} \right)^2 \left( \dfrac{D}{1.79~\kpc} \right)^{-5} \left( \dfrac{\theta}{0.5~\deg} \right)^{-5}.
  \end{equation}
 Substituting Eq.~\eqref{eq: density} for $n$ in Eq.~\eqref{eq: flux} and
  solving for $W_\mathrm{p}$ yields
\begin{equation}
\label{eq: wp_flux}
\begin{aligned}
W_\mathrm{p} = 4.5 \times 10^{49}~\erg & \left( \dfrac{t_\age}{1~\kyr} \right)^{-2} \left( \dfrac{\mathcal{F}_\gamma}{3.9\times10^{-13}~\ph~\cm^{-2}\s^{-1}}\right)  \\ 
& \times \left( \dfrac{D}{1.79~\kpc} \right)^{7}  \left( \dfrac{\theta}{0.5~\deg} \right)^{5}.
\end{aligned}
\end{equation}
Here, we use the observed gamma-ray flux of SN1006,
$\mathcal{F}_\gamma(>1~\rmn{TeV})\approx 3.9 \times10^{-13}~\ph~\cm^{-2}\s^{-1}$,
the angle it subtends over the sky, $\theta\approx0.5$, and the canonical energy
of a SNR, $E_\mathrm{SN} = 10^{51}~\erg$. Estimates for the distance range from
$1.45~\kpc$, calculated using the SNR peak brightness, to $2.2~\kpc$, based on
the comparison of the optical proper motion with the shock velocity derived from
optical thermal line broadening \citep{2003ApJ...585..324W}. More recently,
\citet{2017hsn..book...63K} derived a distance of $1.57\pm 0.07~\kpc$ combining
the shock speed with the proper motion of the North-Western filament. We chose to adopt
an intermediate distance of 1.79~kpc for our model.

\subsubsection{Vela Junior}
The unknown age of Vela Junior increases the uncertainty for the parameter
estimates of Vela Junior in comparison to SN1006.  Estimates on the age vary
from a very young remnant of $\sim 700$ yrs \citep{1999A&A...350..997A} to an
older object of more than 4000~yrs \citep{2008ApJ...678L..35K}.  Distance
estimates are also uncertain. The SNR can be a nearby object at $D = 0.2~\kpc$,
as inferred from studies of the decay of ${}^{44}\mathrm{Ti}$ nuclei
\citep{1998Natur.396..142I}, or a more distant one at $D = 0.75~\kpc$, as
inferred from the slow expansion of X-ray filaments \citep{2008ApJ...678L..35K}.

Regarding the density estimates, the lack of thermal X-ray emission places a
very low limit at $n=0.03~\cm^{-3}$ while assuming a homogeneous environmental
density \citep{2001ApJ...548..814S}. However the interaction with dense clumps
lowers the resulting thermal X-ray emission and allows a higher average
density. A conventional approach in the hadronic model is to use a density of
the order of $n \sim 1~\cm^{-3}$ \citep{2006A&A...449..223A}, while hydrodynamic
models suggest values of less than $0.4~\cm^{-3}$ \citep{2015ApJ...798...82A}.
More recently HI and CO measurements even suggest an extremely high average ISM
density of the order of $n \sim 100~\cm^{-3}$ \citep{2017ApJ...850...71F}.

Adopting the observed flux above $1~\TeV$ for Vela Jr.\ of $2.3 \times 10^{-11}
\ph~\cm^{-2}~\s^{-1}$ \citep{2018AA...612A...7H}, we obtain for $W_\mathrm{p}$
(using Eqs.~\eqref{eq: flux} and \eqref{eq: wp_flux}, respectively):
\begin{align}
W_\mathrm{p} &= 4.5 \times 10^{49}~\erg \left( \dfrac{n}{0.5~\cm^{-3}}\right)^{-1} \left(\dfrac{D}{500~\pc} \right)^2
\nonumber\\
&= 4.5 \times 10^{49}~\erg \left( \dfrac{t_\age}{2.9~\kyr} \right)^{-2}
\left( \dfrac{D}{500~\pc} \right)^{7}  \left( \dfrac{\theta}{2~\deg} \right)^{5}.
\label{eq: wp2}
\end{align}
For an explosion energy of $10^{51}~\erg$, the efficiency adopted by
\citet{2017ApJ...850...71F} only amounts to $\sim 0.1\%$ (i.e. $W_\mathrm{p}=
10^{48}~\erg$) at $D=750~\pc$. Although this low efficiency is compensated by an
extremely high ISM density, in order to maintain a fixed angular size in the sky
at $750~\pc$, from Eq.~\eqref{eq: wp2} we notice that the age of the remnant
would exceed $100$ kyr, far beyond the observational estimates, even for the
extreme case discussed in \citet{2009APh....31..431T}.  The choice for the
distance is determined by the constraints on the age. An SNR age in the range
$(680 - 5100)~\yr$ corresponds to distance ranging between $0.3~\kpc$ and
$0.6~\kpc$ and a density ranging from $0.2~\cm^{-3}$ to $0.66~\cm^{-3}$.
Following the recent estimates on the distance reported in
  \citet{2015ApJ...798...82A} we decided to place the remnant at $D=0.5~\kpc$,
  which would correspond to an age of $2900$ yrs, assuming that the expansion is
  solely in the Sedov-Taylor stage and in a uniform medium.
  
However, a more accurate modeling of the evolution of a core collapse SNR
includes an initial phase in which the SNR is freely expanding with constant
velocity in a wind-blown environment driven only by the initial kinetic energy
and the ejected mass $M_\ej$. Consequently the resulting radius is larger than
the value obtained via Eq.~\eqref{eq: sedov}. Following \citet{1999ApJS..120..299T},
the expanding shock radius that
combines the radii in the free expansion ($r_\free$) and Sedov-Taylor phases
reads:
\begin{equation}
\label{eq: total_radius}
\begin{aligned}
  r_s(t) &= [(r^\ast_{\ST})^{5/2} + r^{5/2}_{\free}]^{2/5} \\
 & = \left[\left(\dfrac{E_\mathrm{SN}}{\alpha \rho}\right)^{1/2} (t_\age-t_\ST) + \left( \dfrac{3 M_\ej}{4\pi \rho}\right)^{5/6} \right]^{2/5}, 
   \end{aligned}
\end{equation}
where $r^\ast_{\ST}$ is the modified Sedov radius starting at the end of the
free expansion phase, and
\begin{equation}
    t_\ST = \dfrac{1}{2} \left( \dfrac{9}{2\pi^2} \right)^{1/6} M_\ej^{5/6} E_\SN^{-1/2} \rho^{-1/3}
    \label{eq:t_ST}
\end{equation}
is the transition time between the free expansion and the Sedov-Taylor regime
corresponding to the moment when the mass swept up by the explosion equals the
ejected mass \citep{1999ApJS..120..299T}.  Combining Eqs.~\eqref{eq: angular}
and \eqref{eq: total_radius} and solving for $t_\age$ we find a more precise
estimate for the age of Vela Jr.

In order to reliably model the circum-stellar medium of Vela Jr., we include
  a population of dense gaseous clumps
  \citep{2018ApJ...866...76M} with a typical size of 0.1~pc and a number density
  of $\sim 10^{3} \cm^{-3}$ \citep{2012ApJ...744...71I}.  The detection of these
  molecular clouds is linked to the rotational CO lines often observed in these
  systems \citep{2013Fukui}. However, it is questionable whether future
  telescopes will have enough resolution to resolve the emission from a single
  clump. To include the effect of the clumpy ISM, we redefine Eq.~\eqref{eq:
    flux} as:
\begin{equation}
\label{eq: flux clumps}
\begin{aligned}
\mathcal{F}_\gamma \simeq & ~2.7 \times 10^{-12} (1 +\chi)   \\
& \times \left( \dfrac{W_\mathrm{p}}{10^{50}\,\erg}\right)
  \left( \dfrac{n}{0.1\,\cm^{-3}}\right) \left( \dfrac{D}{1\,\kpc} \right)^{-2}\!\! \dfrac{\ph}{\cm^{2}~\s},
 \end{aligned}
\end{equation}
where $\chi$ is the ratio between the swept-up mass of the clumps within the SNR volume $V_{\SNR}$ and the diffuse ISM mass swept up by the shock. It reads:
\begin{equation}
\label{eq: chi factor}
\begin{aligned}
\chi &= \bar{\eta} \dfrac{M^{\swept}_\mathrm{c}}{M^{\swept}_{\mathrm{ISM}}} = \bar{\eta} \dfrac{\langle \rho_{\mathrm{c}}\rangle V_{\SNR}}{\rho_{\mathrm{ISM}} V_{\SNR}} \\ 
&= 0.2 \left( \dfrac{\bar{\eta}(t)}{30\%}\right) \left( \dfrac{\langle\rho_{\mathrm{c}}\rangle}{1.6 \times 10^{-2}~M_\odot \pc^{-3}}\right) \left( \dfrac{n}{0.42~\cm^{-3}}\right)^{-1},
\end{aligned}
\end{equation}
where $\bar{\eta}(t) = 30\%$ is the average percentage of clumped mass penetrated
  and accelerated by the shock and $\langle \rho_\mathrm{c} \rangle$ is the
  average density of dense gas contained in the clumps. More details
  about the behavior of $\bar{\eta}(t)$ can be found in Appendix~\ref{sec: appx}.
  Inserting Eq.~\eqref{eq: chi factor} into Eq.~\eqref{eq: flux clumps} and
  solving for the ISM density $n$, we find:
\begin{equation}
\label{eq: density and clumps}
\begin{aligned}
n = 0.42~\cm^{-3} &\biggl[  1.2 \left( \dfrac{\mathcal{F}_\gamma}{2.3 \times 10^{-11}~\ph~\cm^{-2}~\s^{-1} } \right) \\
& \times   \left( \dfrac{W_\mathrm{p}}{4.5 \times 10^{49}~\erg}\right)^{-1}  \left(\dfrac{D}{0.5~\kpc}\right)^2 \\
 &- 0.2 \left( \dfrac{\bar{\eta}(t)}{30\%}\right) \left( \dfrac{\langle\rho_{\mathrm{c}}\rangle}{1.6 \times 10^{-2}~M_\odot \pc^{-3}}\right) \biggl] .
\end{aligned}
\end{equation}
In our setup for Vela Jr. we set $\langle\rho_{\mathrm{c}}\rangle = 1.6 \times 10^{-2}~M_\odot \pc^{-3}$ and we assume a distance of $D=0.5$~kpc so that the dense-cloud mass in within the supernova remnant volume is similar to the value assumed by \citet{2019MNRAS.487.3199C} for the cloud mass in RX-J1713,
\begin{equation}
\begin{aligned}
M^{\swept}_\mathrm{c} = \dfrac{4}{3} \pi r^3_s \langle \rho_\mathrm{c} \rangle =& 45 M_\odot \left( \dfrac{D}{0.5~\kpc}\right)^3 \left( \dfrac{\theta}{2~\deg}\right)^3  \\ 
& \times \left( \dfrac{ \langle \rho_\mathrm{c} \rangle}{1.6 \times 10^{-2}~M_\odot \pc^{-3}}\right) 
\end{aligned}
\end{equation}
where we expressed $r_s$ by Eq.~\eqref{eq: angular}. 

Thus, the corresponding diffuse inter-clump density for the ISM is lowered to
  $ n =0.42~\cm^{-3}$ (see Eq.~\ref{eq: density and clumps}) and the
  corresponding age as a function of the ejected mass is
\begin{equation}
  t_\age = 2.68~\kyr  +0.12~\kyr \left( \dfrac{M_\ej}{M_\odot} \right)^{5/6}.
\end{equation}
If we assume an ejected mass of $3~M_\odot$, the resulting age for Vela Jr.\ is
$3000$ yrs, slightly higher than that inferred from considering a
Sedov-Taylor phase only in a non-clumpy medium. We adopt this more
accurate estimate in our analysis re-scale the distances of our Sedov-Taylor
simulations according to Eq.~\eqref{eq: total_radius}. We defer a detailed
simulation of this combined free expansion and Sedov-Taylor phases to future
work, since the dominating CR pressure inside the remnant should also cause the
Rayleigh-Taylor instabilities at the contact discontinuity of shocked ISM and
shocked ejecta to develop differently.

\begin{table*}
\caption{Comparison of simulation and observational parameters}
\begin{center}
\begin{tabular}{c||c|c|c|c|c|c}
\hline
\textbf{SNR} & \multicolumn{3}{|c|}{\textbf{SN1006}} & \multicolumn{3}{|c|}{\textbf{Vela Jr.}}  \\ 
\hline
\hline
Parameter & Simulation &  Observation & Reference & Simulation & Observation & Reference \\ 
\hline
diameter $\theta_{\rmn{s}}$ [deg] & 0.5 & 0.5 & 7 & 2 & 2 & 7 \\ 
spectral index $\Gamma_\mathrm{HE}$ & 1.95 & $ 1.79 \pm 0.44 $ & 2, 8  & 1.81 & $1.85 \pm 0.24 $ & 14 \\ 
spectral index $\Gamma_\mathrm{TeV}$ & 2.15 & $2.30 \pm 0.15$ & 2, 8 & 2.11 & $2.24 \pm 0.19 $ & 2 \\ 
$\mathcal{F}_\gamma(>1\mathrm{TeV})~[10^{-12}\mathrm{ph}~\mathrm{cm}^{-2}~\mathrm{s}^{-1}]$  & $0.39$ & $0.39 \pm 0.08$ & 9 & $23.4$ & $ 23.4 \pm 5.6 $  & 9  \\ 
\hline
$n$ $[\mathrm{cm}^{-3}]$ & $0.1^\mathbf{H}$, $0.07^\mathbf{M}$  & $0.05 - 0.3$  & 1, 5 & $0.42^\mathbf{H}$, $0.38^\mathbf{M}$  & $0.03 - 100 $ & 3, 6, 13 \\ 
$M_c~[M_\odot]$  & $-$  & $-$  & $-$ & $45^\mathbf{H, M}$  & $-$ & $-$ \\ 
$D$ [pc]  & $1790^\mathbf{H}$, $1930^\mathbf{M}$ & $1450-2200$ & 2, 11 & $500^\mathbf{H}$, $500^\mathbf{M}$ & $200-750$ & 4, 10  \\ 
diameter $d_{\rmn{s}}$ [pc] & $13.3^\mathbf{H}$, $14.3^\mathbf{M}$ & $12.6-19.2$ & $-$ & $17.3^\mathbf{H}$, $17.3^\mathbf{M}$ & $7-26$ & $-$  \\ 
$t_\age$ [yrs] & 1012 & 1012 & $-$  & $3000^\mathbf{H}$ ,  $2860^\mathbf{M}$ & $680 - 5100$ & 4, 10, 12 \\ 
$v$ [km s$^{-1}$] & $3000^\mathbf{H}$ , $3200^\mathbf{M}$ & 2100 - 4980 & 1, 11, 15 & $1800^\mathbf{H}, 2000^{\mathbf{M}}$ & $> 1000$ & 3 \\
$B$ [$\muG$] & $200^\mathbf{H}, 80^\mathbf{M} $& $-$ & $-$ &  $40^\mathbf{H}, 10^\mathbf{M} $ & $-$ & $-$ \\
\hline
$E_{\mathrm{p}, \mathrm{cut}} [\TeV]$ & 200 & $-$ & $-$ & 100 & $-$& $-$ \\
$E_{\mathrm{e}, \mathrm{cut}} [\TeV]$ & $1.7^\mathbf{H}$,  $3^\mathbf{M}$ & $-$& $-$ &  $0.25^\mathbf{H}$,  $0.4^\mathbf{M}$ & $-$ & $-$ \\
$\beta_\mathrm{p}$ & 2 & $-$ &$-$ & 2 & $-$ & $-$ \\
$\beta_\mathrm{e}$ & 0.7 & $-$ &$-$ & 0.4 & $-$ & $-$ \\
\hline
\end{tabular} \\[.5em]
\end{center}
Notes: $n$ denotes the diffuse ISM number density, 
$M_\mathrm{c}$ is the target clump mass hit by the remnant, $D$ is the distance to the SNR,
$\theta_{\rmn{s}}$ and $d_{\rmn{s}}$ are the angular and proper extent of the
blast wave, $v$ denotes the shock velocity, $t_\age$ is the SNR age, $B$ the magnetic
field and $\mathcal{F}_\gamma$ is the gamma-ray flux. The last four lines
represent the parameters used for the spectral models described in
Sec.~\ref{sec: Multifreq}. The superscripts \textbf{H} and \textbf{M} denote the
parameters of the pure hadronic and mixed hadronic/leptonic models,
respectively.\\
References:  (1) \cite{2007AA...475..883A} ; (2) \cite{2015AA...580A..74A}; (3) \cite{2015ApJ...798...82A}; (4) \cite{1999A&A...350..997A}; (5) \cite{2002A&A...387.1047D}; (6) \cite{2017ApJ...850...71F}; (7) \cite{2014BASI...42...47G}; (8) \cite{2010AA...516A..62A};  (9) \cite{2018AA...612A...7H}; (10) \cite{2008ApJ...678L..35K}; (11) \cite{2017hsn..book...63K}; (12) \cite{2019PhRvD.100b4063M}; (13) \cite{2001ApJ...548..814S}; (14) \cite{2011ApJ...740L..51T}; (15) \cite{2003ApJ...585..324W}.
\label{table:Table 1}
\end{table*}

\begin{figure*}
  \begin{center}
    \includegraphics[width=\textwidth]{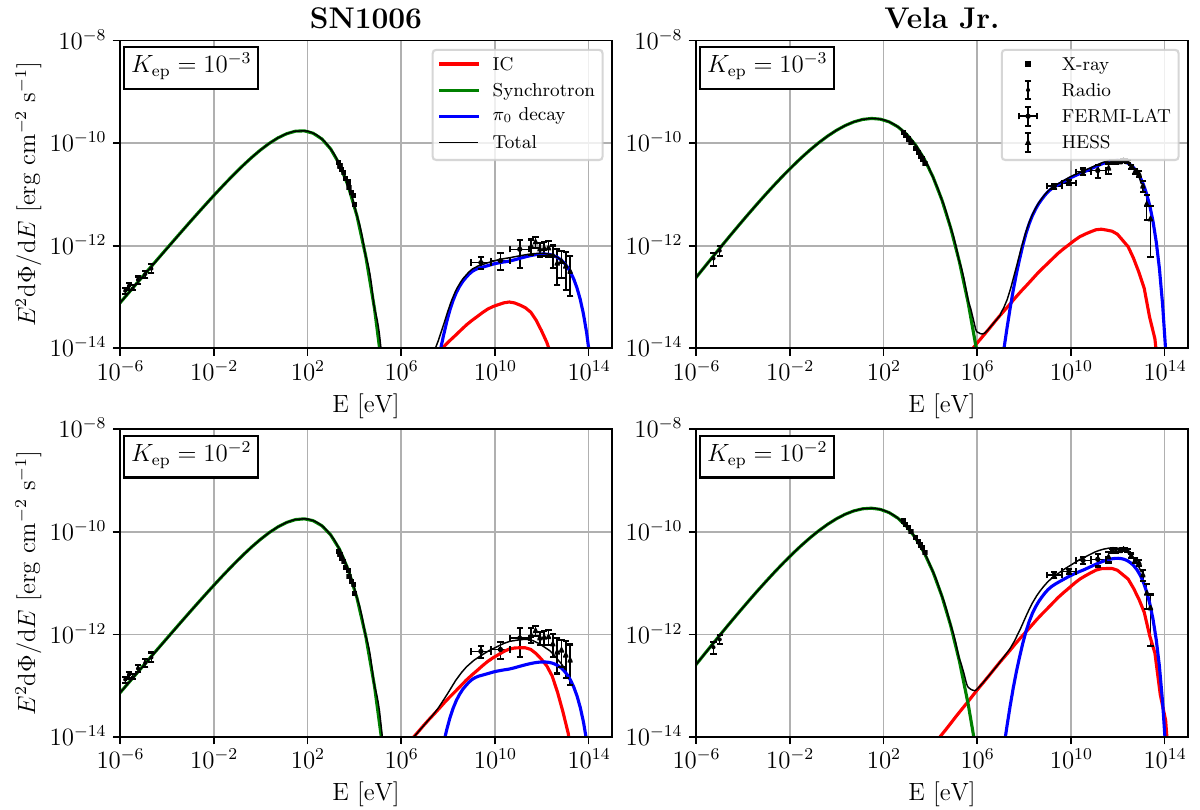}
  \end{center}
  \caption{Multi-frequency spectra of SN1006 (left-hand panels) and Vela
    Jr.\ (right-hand panels). The top panels show a hadronic scenario for both
    remnants assuming an electron-to-proton ratio of $K_{\rmn{ep}}=10^{-3}$. The
    bottom panels show a mixed hadronic-leptonic scenario with
    $K_{\rmn{ep}}=10^{-2}$. For SN1006, we use data in the radio
    \citep{1996ApJ...459L..13R}, X-rays \citep{2008PASJ...60S.153B}, from FERMI
    \citep{2010ApJS..188..405A} and H.E.S.S.\ \citep{2010AA...516A..62A} (sum of
    the two regions). For Vela Jr., we adopt data in the radio
    \citep{2000A&A...364..732D}, X-rays \citep{2007ApJ...661..236A}, from FERMI
    \citep{2011ApJ...740L..51T} and H.E.S.S.\ \citep{2018AA...612A...7H}. We
    account for the following processes: synchrotron radiation from primary
    electrons (green lines), IC scattering on the CMB for SN1006 and
    additionally on starlight for Vela Jr.\ (red lines) and hadronic
    interactions (blue lines).}
  \label{fig:multifrequency}
\end{figure*}

\section{Multi-frequency spectral modelling}
\label{sec: Multifreq}

In order to improve the order of magnitude limits as presented in
Sect.~\ref{sec:obs}, we derive multi-wavelength spectra from radio to TeV
gamma-rays. This enables us to
carefully investigate the nature of the gamma-ray emission from both SNRs.  The
data are then compared to a one-zone model in which the integrated particle
populations (electrons and protons, denoted by subscripts $i = \lbrace
\mathrm{e}, \mathrm{p} \rbrace$) are described by a power law with exponential
cutoff of the form:
\begin{equation}
f^{\rmn{1D}}(p_i) = \dfrac{\de^2 N_i}{\de p_i \de V} \propto p_i^{-\alpha_i} \exp\left[ -\left( \dfrac{p_i}{p_{i,\mathrm{cut}}} \right)^{\beta_i} \right]
\end{equation}
where $f^{\rmn{1D}}(p_i)=4\pi\,p_i^2f^{\rmn{3D}}(p_i)$, $\alpha_i$ is the
spectral index, $p_{i,\mathrm{cut}}$ is the cutoff momentum and $\beta_i$
describes the sharpness of the cutoff; with values reported in
Table~\ref{table:Table 1}.

\begin{figure*}
  \begin{center}
    \includegraphics[width=\textwidth]{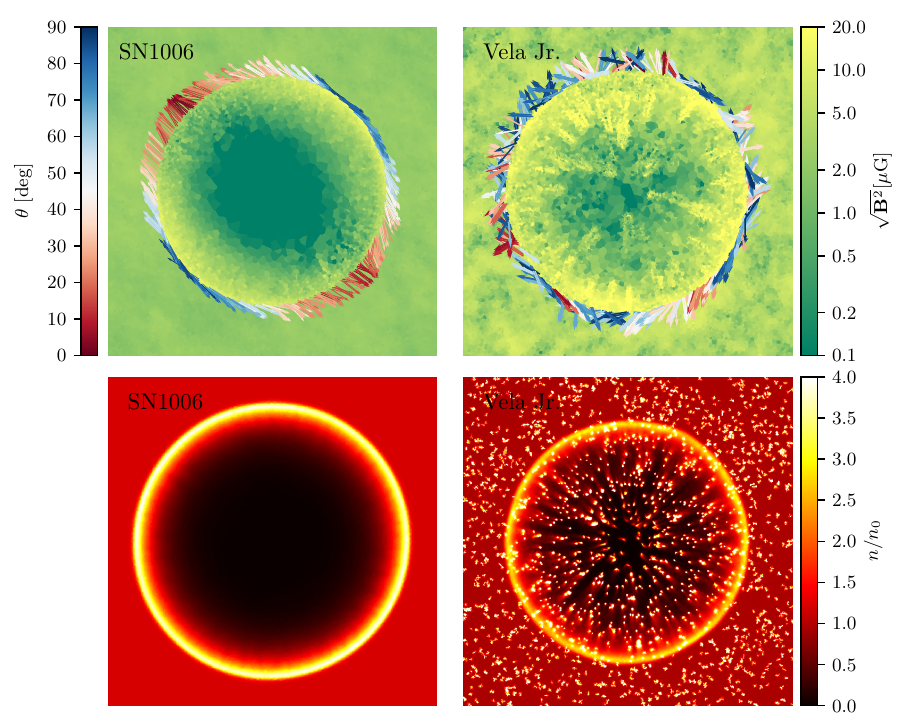}
  \end{center}
  \caption{Two-dimensional cross-sections of our SNR simulations.
    Top row: slices of the magnetic field strengths of SN1006 and Vela
    Jr.\. Bottom row: 2D slices of the number density $n$ normalized to the background density $n_0$.  
    The outwards pointing arrows in the
    top panels show the orientation of the magnetic field at the shock,
    color-coded by the magnetic obliquity (red/blue for
    quasi-parallel/-perpendicular shocks). Both simulation models adopt a
    constant-density ISM and differ only in the assumed magnetic morphology:
    SN1006 has a homogeneous magnetic field pointing to the top-left augmented
    with a mildly turbulent field while Vela Jr.\ adopts a fully turbulent
    magnetic field with correlation length $\lambda_B=L/2=13$~pc. }
  \label{fig:comparison1}
\end{figure*}

\begin{figure*}
  \begin{center}
    \includegraphics[width=\textwidth]{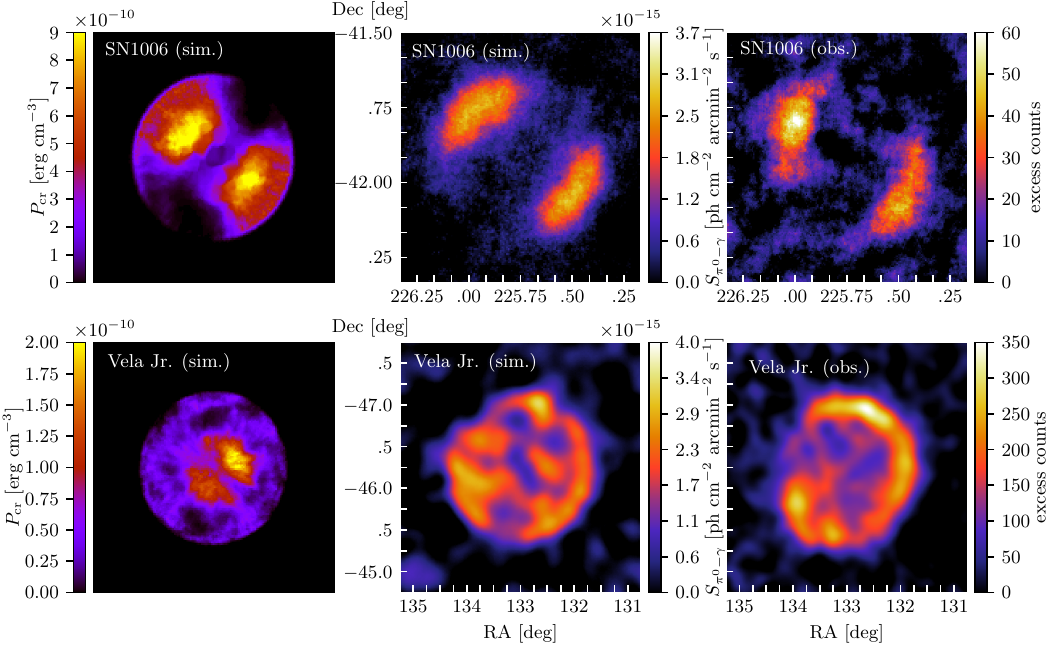}
  \end{center}
  \caption{Two-dimensional projected gamma-ray maps of our SNR simulations. Top
    row, featuring SN1006: CR pressure (left), simulated pion-decay gamma-ray
    surface brightness resulting from hadronic CR interactions convolved to the
    observational resolution (middle) and acceptance-corrected excess map for
    SN1006 with a Gaussian PSF of width $R_{68} = 0.064^\circ$, 
    corresponding to $\sigma = R_{68}/1.515  = 0.042^\circ$ 
    \citep{2010AA...516A..62A}.  Bottom row, featuring the same quantities for
    Vela Jr.\ with an acceptance-corrected excess map that was convolved with a
    Gaussian PSF of width $\sigma = 0.08^\circ$ \citep{2018AA...612A...7H}.  We
    add Gaussian noise at the observed level and power spectrum and convolve
    both simulated gamma-ray maps to the observational angular resolution.}
  \label{fig:comparison2}
\end{figure*}

Radio synchrotron and inverse Compton emission (including the Klein-Nishina
cross section) are calculated following \citet{1970RvMP...42..237B}. The
hadronic gamma-ray emission is calculated from parametrisations of the
cross-section of neutral pion production at low and high proton energies,
respectively \citep{2018arXiv180305072Y,2006PhRvD..74c4018K}. Because our
simulations only follow CR protons, we assume an electron-to-proton ratio
$K_{\mathrm{ep}}$ at 10~GeV for the normalization of the electron
population. For our hadronic model we set $K_{\mathrm{ep}} = 10^{-3}$.  Since
there are parameter degeneracies, we also show a mixed hadronic/leptonic model
for the gamma-ray emission with $K_{\mathrm{ep}} = 10^{-2}$.

Results for the pure hadronic and the mixed hadronic-leptonic models are shown
in Fig.~\ref{fig:multifrequency} for both SNRs. The model parameters are
reported in the lower section of Tab.~\ref{table:Table 1}. Note that the
magnetic field entering here is the radio synchrotron emission-weighted magnetic
field, which is situated in the post-shock region, interior to the SNR shell.
In the case of SN1006, because of its position above the galactic plane, we
assume an inverse-Compton (IC) scattering mainly on CMB photons. The location of
Vela Jr.\ in a star forming region suggests that a combination of IC scattering
on starlight with an energy density of $5 u_{\mathrm{CMB}}$ and CMB photons is
more appropriate. In particular, we assume that the starlight is reprocessed
  by warm dust with a temperature of 100 K, which is typical for conditions in
  star forming regions \citep{2012A&A...538A..81M}.  Note that we adopt hard CR
proton spectral indices of $\alpha_\rmn{p}<2$ in all models, which naturally
emerge as a result of streaming CRs inside dense clumps of a clumpy ISM
\citep{2019MNRAS.487.3199C}.

\section{Morphological gamma-ray modelling}
\label{sec:morphological_modelling}

\begin{figure*}
  \begin{center}
    \includegraphics[width=1.0\textwidth]{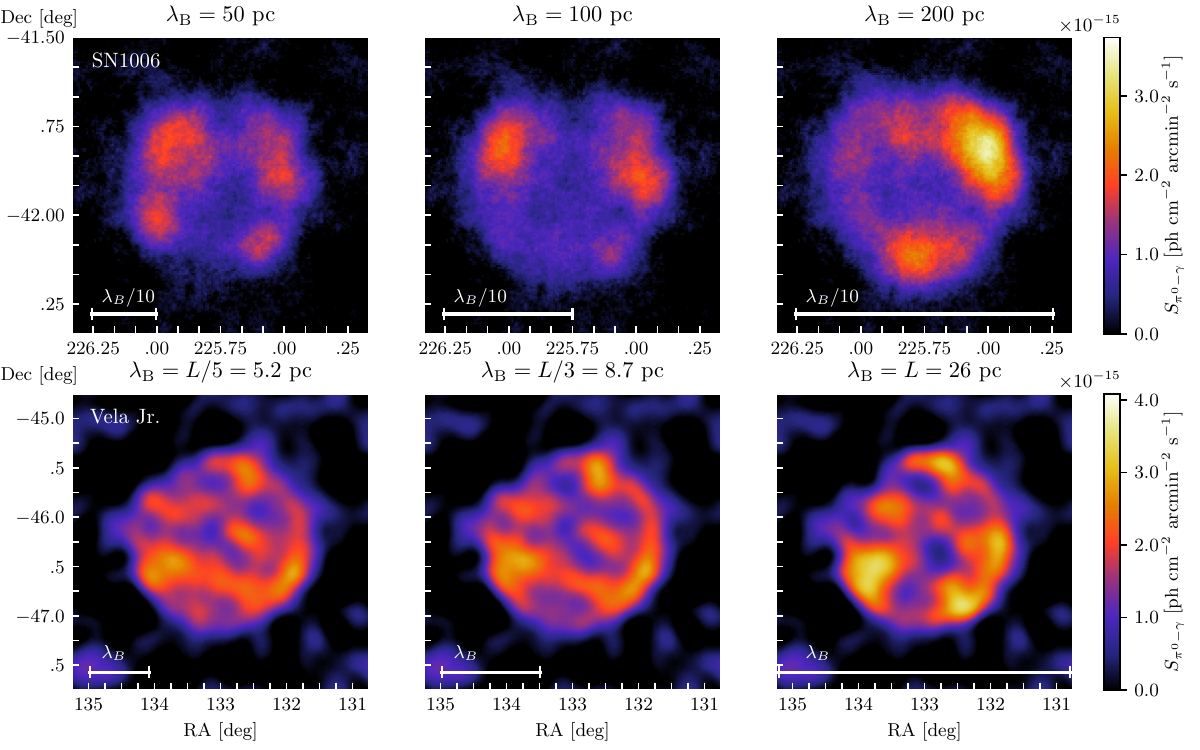}
  \end{center}
  \caption{Synthetic gamma-ray maps of SN1006 (first row) and Vela
    Jr.\ (second row) for a purely turbulent magnetic field with different
    coherence lengths (see panels). The sequence towards larger correlation
    lengths starts to approach more homogeneous magnetic field geometries with
    the characteristic bi-lobed shell morphology (top right for SN1006) whereas
    smaller coherence lengths approach the isotropic limit (lower left for Vela
    Jr.).  Clearly, the observed gamma-ray map of Vela Jr.\ falls in between the
    cases $\lambda_B=8.7$ and $26$~pc (bottom left and middle panels),
    justifying our choice of $\lambda_B=13$~pc.}
  \label{fig:correlation}
\end{figure*}

Our simulation models for the two SNRs are described by the
energy-conserving Sedov-Taylor solution \citep{1988RvMP...60....1O}. The
solution also remains self similar for obliquity-dependent CR acceleration
\citep{2018MNRAS.478.5278P}.  We assume a power-law CR momentum distribution for
calculating the pion-decay gamma-ray emissivity
\citep{2004A&A...413...17P,2008MNRAS.385.1211P}.  After line-of-sight
integrating the gamma-ray emissivity and adding Gaussian noise (so that the
synthetic map matches the observational noise properties in amplitude and scale,
see Pais \& Pfrommer in prep.), we convolve the maps with the observational
point-spread function.

First, we perform exploratory simulations with parameter choices guided by the
self-similar scaling of the Sedov-Taylor solution. We find parameter
combinations that approximately reproduce all observational characteristics
(with box size $L = 20$~pc for SN1006 and $L = 26$~pc for Vela Jr.). Fixing
angular size, explosion energy, and employing the self-similar solution, we then
re-scale the solution by varying the ambient density within observational bounds
to match the observed gamma-ray fluxes.  In case of the core collapse
  supernova remnant Vela Jr.\ we must take into account the free expansion
  phase, while assuming an ejected mass of $3~\rmn{M}_\odot$. Thus, we evolve
  the simulation of the Sedov explosion so that the combination of free
  expansion and Sedov phase matches its angular size at a given distance following
  Eq.~\eqref{eq: total_radius}, and we re-scale the shock radius accordingly to
  account for the free expansion phase.  The final set of parameters is
reported in Table~\ref{table:Table 1}.

To model SN1006 we assume a dominant homogeneous magnetic field that points to
the top-left as supported by studies of radio polarization signatures
\citep{2013AJ....145..104R}. We superpose a turbulent magnetic field with a
correlation length $\lambda_B=L=20~\pc$ ($\approx 0.74^{\circ}$ at $D=1.53~\kpc$)
that contains 1/9 of the energy density of the homogeneous field. For Vela
Jr.\ we perform a range of fully turbulent simulations with magnetic coherence
lengths $\lambda_B=L/f$ ($f\in\{1,2,3,4,5\}$). We find that our simulation model
with $\lambda_B=L/2=13$~pc ($\approx 2.3^\circ$ at $0.5$~kpc) statistically
matches the gamma-ray maps best.

We present different physical properties of our simulation models for SN1006 and
Vela Jr.\ in Fig.~\ref{fig:comparison1} and \ref{fig:comparison2}. While both
simulation models adopt a constant background density, their magnetic structure
differs (Fig.~\ref{fig:comparison1}). This results in a significantly different
CR pressure distribution owing to the obliquity-dependent shock acceleration
(left column of Fig.~\ref{fig:comparison2}).

The hadronically induced gamma-ray maps echo this difference as they depend on
the CR pressure distribution multiplied with the target gas density, which peaks
at the shock surface (middle column of Fig.~\ref{fig:comparison2}). The bi-lobed
gamma-ray morphology of SN1006 is a direct consequence of quasi-parallel shock
configuration at the polar caps. This contrasts with the patchy filamentary,
limb-brightened gamma-ray morphology of our model for Vela Jr., which results
from the small-scale coherent magnetic patches with a quasi-parallel shock
geometry.

A direct comparison with observational images is shown in the right column of
Fig.~\ref{fig:comparison2}. In the case of SN1006, we convolve the simulated map
with a Gaussian of width $\sigma=0.042^\circ$ (equal to $\sigma  = R_{68}/1.515$ where $R_{68} = 0.064^\circ$), in the
case of Vela Jr.\ we use $\sigma=0.08^\circ$ (the observational point spread
function, PSF). The obliquity-dependent shock acceleration model is able to
accurately match the TeV gamma-ray morphologies of SN1006 and Vela
Jr.\ solely by changing the magnetic coherence scale (with a homogeneous field
representing the limit of an infinite coherence scale). Clearly, in the case of
Vela Jr.\ this match is on a statistical basis as the phases of turbulent fields
are random. We emphasize that all our
simulations assumed a constant-density ISM that the SNR has expanded into. Note
that we also obtain filamentary gamma-ray morphologies due to obliquity
dependent shock acceleration in SNRs that are expanding into a stellar wind
environment (see Sec.~\ref{sec: stellar wind}).

The success of our models enables us to estimate $\lambda_B$ of the ISM
surrounding SN1006 and Vela Jr.\ by comparing the observed gamma-ray maps to
simulations with different values of $\lambda_B$. While the morphology of SN1006
is best matched by a homogeneous ambient field (possibly with the addition of a
small-amplitude turbulent field), we need to perform an analysis similar to Vela
Jr.\ in order to formally place a lower limit on the magnetic coherence
length. To this end, we perform three different simulations that have a purely
turbulent field with coherence scales of $\lambda_B=50$, 100 and 200~pc.
Figure~\ref{fig:correlation} shows gamma-ray maps of three different magnetic
coherence scales for both SNRs, respectively. For SN1006, the number of
gamma-ray patches decreases with increasing coherence scale (left to right) to
the point where there are two patches visible ($\lambda_B=200~\pc$). Since the
alignment of these two patches is not symmetric with respect to the centre, we
conclude that the true coherence scale must be larger and in fact is consistent
with a nearly homogeneous field across the SNR. For Vela Jr., the sequence of
gamma-ray maps with decreasing coherence scale leads to smaller-scale gamma-ray
patches that asymptotically approach an isotropic distribution. We find that the
correlation length of Vela Jr.\ ranges in between the box size $L$ and $L/3$,
suggesting $\lambda_B\approx L/2=13_{-4.3}^{+13}$~pc, allowing for
uncertainties in distance and $\lambda_B$.  

\section{SNR expanding into a stellar wind}
\label{sec: stellar wind}

\begin{figure*}
  \begin{center}
    \includegraphics[width=1.0\textwidth]{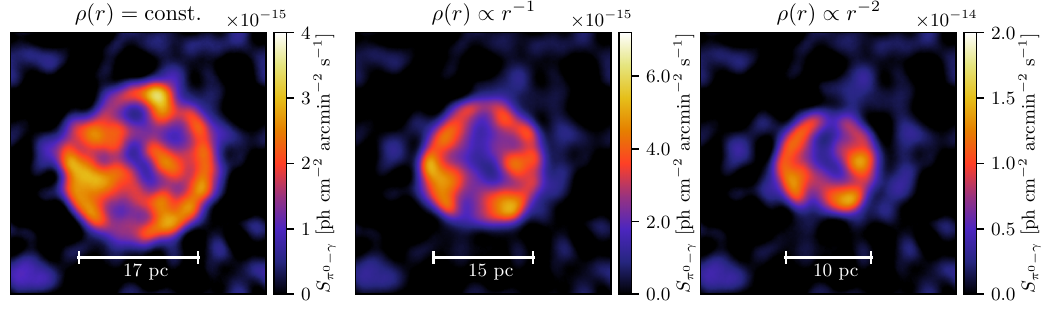}
  \end{center}
  \caption{Synthetic $\gamma$-ray maps of Vela Jr., which result from
    simulations of a blast wave expanding into a stellar wind profile. To
    account for the uncertainties, we adopt different wind density profiles and
    a purely turbulent magnetic field of coherence scale $\lambda_B=13$~pc. The
    panels show SNRs that propagate in a constant density medium (left), in a
    density profile $\rho\propto r^{-1}$ (middle) and $\rho\propto r^{-2}$
    (right) at the same age and same mean density. As expected, the higher
    central density slows down the expanding shock, modifies the morphology into
    a more compact emission region, which however maintains the shell-type
    morphology.}
  \label{fig:winds}
\end{figure*}

Here, we study how different assumptions of the circumstellar medium (CSM)
affect the evolution of a SNR and its morphological appearance at gamma-ray
energies.  Studies of SNR evolution into stellar-wind-blown environments range
from the initial free-expansion phase \citep{2010ApJ...725..922S,
  2014ApJ...797....2K, 2015ApJ...806L..19F} to the self-similar Sedov phase
\citep{1999ApJ...527..866L}.

The Vela Jr.\ SNR is thought to be associated with a core-collapse supernova
explosion \citep{2002ApJ...574..155W}, which results from the collapse of a
massive star of mass $M \geq 8 M_\odot$.
The particularities of the progenitor are
responsible for the evolution of the SNR in a highly modified wind-blown CSM
shell, causing a substantially different evolution from the classical sequence
of free expansion followed by a Sedov and a radiative stage
\citep{2005ApJ...630..892D}. As pointed out by \citet{1982ApJ...258..790C} and
\citet{1988RvMP...60....1O}, a SNR that interacts with a CSM density profile
$\rho(r) \propto r^{-q}$ has a self-similar analytical solution for the
evolution of shock radius and velocity:
\begin{align}
r_\rmn{s}(t) &=  A^{\frac{1}{5-q}} t^{\frac{2}{5-q}},\\
v_\rmn{s}(t) &=  \dfrac{2}{5-q} A^{\frac{1}{5-q}} t^{\frac{q-3}{5-q}},
\end{align}
where $A$ is a constant depending on the ambient average density, the SN energy
and the adiabatic index. Here, $q=2$ corresponds to the case of constant mass
loss from the progenitor star.

We simulate supernova explosions in three different power-law wind profiles with
$q\in\{0,1,2\}$. To ease comparison we adopt the same average number density
for all simulations as reported in Table~\ref{table:Table 1}.  All other initial
simulation parameters for the energy and the turbulent magnetic field remain
unchanged.  We evaluate the SNR simulations at the same age, which emphasizes
the effect of a denser central CSM for steeper power-law indices. In order to
avoid a non-vanishing magnetic divergence during the generation of the initial
density profile, we cap the density in the central cells with a Plummer-type
softening length of $r_\rmn{c}=0.3$~pc.

The wind speed $v_w$ ranges from values of order $(1000-3000)~\km\ \s^{-1}$ for
very young SNRs \citep{1978ApJ...225..893A} to $100\ \km\ \s^{-1}$ or less for
red-giant stars. Hence, we can neglect $v_w$ in comparison to the shock velocity
during the early Sedov phase \citep{1988RvMP...60....1O}, which means that the
approximation of assuming a point explosion in the various density profiles is
fully justified and does not affect the final simulation result. This causes the
remnant to directly enter the Sedov stage and to bypass the earlier phase of a
swept-up wind-blown shell. 

In Fig.~\ref{fig:winds} we show simulated gamma-ray maps of Vela Jr.\ for the
three different CSM environments. The primary effect of a stratified wind
density profile consists of slowing down the propagating blast wave. This
results in a more compact and brighter gamma-ray morphology. Many of the general
morphological features of the patchy gamma-ray map previously found for the
constant density solution carry over to the stratified density
profiles. However, the higher central number density in comparison to the flat
profile increases the gamma-ray brightness, with a flux enhancement by a factor
of six for the $q=2$ profile, as expected for the evolution of these profiles
at early times \citep{1995A&A...293L..37K}. Thus, comparing our simulated maps
to the observed shell-type morphologies at these ages, this argues for more
shallow density profiles $|q|<1$, with slight preferences for a constant density
medium for the SNRs studied here.

\section{Discussion and Conclusions}
\label{sec:conclusion}
We have presented the first global simulations and gamma-ray maps of SNRs in
  the hadronic model, which account for magnetic obliquity-dependent CR
acceleration.  We show that the multi-frequency spectrum in the hadronic and
  mixed hadronic/leptonic models match observational data for our simulation
  parameters of the ISM, which are motivated by observations.\footnote{We
      note that the purely leptonic scenario can also match the multi-frequency
      data of our SNRs (SN1006, \citealt{2010AA...516A..62A}; Vela Jr.,
      \citealt{2018AA...612A...7H}). We will address the interesting question
      whether three-dimensional MHD simulations with obliquity dependent
      electron acceleration can produce radio, X-ray and gamma-ray maps in the
      leptonic model that match observational data in future work.}  Our
synthetic gamma-ray maps match the apparently disparate TeV morphologies and
total gamma-ray fluxes of SNR 1006 and Vela Jr.\ within a single physical model
extremely well: SN1006 expands into a homogeneous magnetic field that is
reminiscent of conditions for a galactic outflow or a large-scale Parker loop as
supported by its Galactic height of $z=0.6$ kpc (at $D\simeq 1.8~$kpc) above the
midplane \citep{2002ISAA....5.....S}.  On the contrary, Vela Jr.\ is embedded in
a small-scale turbulent field typical of spiral arms. This suggests that the
diversity of shell-type TeV SNRs originates in the obliquity dependence of the
acceleration process rather than in density inhomogeneities.

Comparing our simulations of different $\lambda_B$ to observed TeV maps of
shell-type SNRs enables us to estimate $\lambda_B$ of the unperturbed ISM before
it encountered the SNR blast wave. Assuming statistical homogeneity, we
constrain $\lambda_B$ in the vicinity of SN1006 and Vela Jr.\ to
$>200_{-10}^{+80}$~pc and $13_{-4.3}^{+13}$~pc, respectively. Simulating the SN
explosion that expands into a stratified density profile caused by a stellar
wind produces similarly patchy gamma-ray maps and hence does not alter our
conclusions that magnetic obliquity-dependent CR acceleration is responsible for
this patchy morphology. However, at the same mean density, the blast wave will
encounter a denser CSM at small radii, which slows down the propagating blast
wave and results in a more compact and brighter gamma-ray map at the same age.

If obliquity-dependent diffusive shock acceleration also applies to electrons,
we could produce similar synthetic TeV maps in the leptonic model to constrain
the magnetic coherence length. If electron acceleration were independent of
magnetic obliquity then this work would provide strong evidence for the hadronic
scenario in shell-type SNRs as the necessary element to explain the patchy TeV
emission. In any case, we conclude that the inferred coherence scales are robust
to specific assumptions of the gamma-ray emission scenario (hadronic
vs.\ leptonic models).

Moreover, here we show that the hadronic model is able to explain shell-type SNR
morphologies, which naturally emerge in our simulations due to the peaked
density at the shock in combination with the slowly decreasing CR pressure
profile \citep{2018MNRAS.478.5278P}. In the leptonic model, fast electron
cooling would have to confine the emission regions close to the shock. However,
this would imply strong spectral softening towards the SNR interior, which is
not observed in Vela Jr., seriously questioning the leptonic model for this SNR
\citep{2018AA...612A...7H}.  Our work opens up the possibility of mapping out
the magnetic coherence scale across the Milky Way and other nearby galaxies at
the locations of TeV shell-type SNRs, and to study how it varies depending on
its vertical height or its location with respect to a spiral arm. Thus, our work
represents an exciting new science case for gamma-ray astronomy, in particular
for the Cherenkov Telescope Array.

\section*{Acknowledgements}
We would like to thank our anonymous referee for a detailed and insightful
report which helped us improve the manuscript.  We would like to thank the
H.E.S.S.\ Collaboration, in particular Fabio Acero, for providing us with the
published data of both SNRs. We acknowledge support by the European Research
Council under ERC-CoG grant CRAGSMAN-646955. This research was supported in part
by the National Science Foundation under Grant No. NSF PHY-1748958.

\bibliographystyle{mnras}
\bibliography{ms}
\newpage
\appendix

\section{Analytical model for the gamma-ray luminosity in a clumped medium}
\label{sec: appx}
Here, we study the interaction of a shock with a dense cloud with numerical
  simuations and analytics and construct an analytical model for the gamma-ray
  luminosity from a core collapse SNR. We assume that the shock propagates
  through a highly structured ISM with a large population of dense clouds, which
  plays an important role for the resulting multi-frequency emission
  \citep{2002ApJ...574..155W}.

\subsection{Numerical setup}
To capture the clumpy structure of the ISM in these regions we generated $ 7
  \times 10^{3}$ dense gaseous clumps with a radius of $R_\mathrm{c} = 0.05~\pc
  $ each and uniformly distributed in a box of size $L = 25~\pc$, subset of the
  simulation box of size $L=26~\pc$.  We used a margin of $0.5~\pc$ for each
  side of the cube to avoid any incidental misplacement of the clumps.  We
  deposit a total mass of dense clumps of $M^{\mathrm{tot}}_{\mathrm{c}} =
  255~M_{\odot}$ within the selected box so that each clump has a mass of
  $M_\mathrm{c} = 0.036~ M_{\odot}$ and the volume-averaged $\mathrm{H}_2$
  density is $\langle \rho_\mathrm{c} \rangle = 1.6 \times 10^{-2}~M_\odot
  \pc^{-3}$. Following a setup similar to \citet{2019MNRAS.487.3199C}, we assume
  spherical clumps with a number density $n_\mathrm{c}=1.4 \times 10^3~\cm^{-3}$
  and a molecular weight $\mu = 2$ (associated to the clumps) so that the density contrast with respect to
  the background is $\delta = n_\mathrm{c} / n_0 \sim 3 \times 10^{3}$. Note
  that we neglect radiative cooling in these simulations, which is negligible
  over the propagation time of the SNR shocks considered while it is crucial for
  understanding the (thermo-)dynamics of the system on longer time scales
  \citep{2018MNRAS.473.5407M,2018MNRAS.480L.111G,2019MNRAS.482.5401S}. 

\begin{figure*}
  \begin{center}
    \includegraphics[scale=0.72]{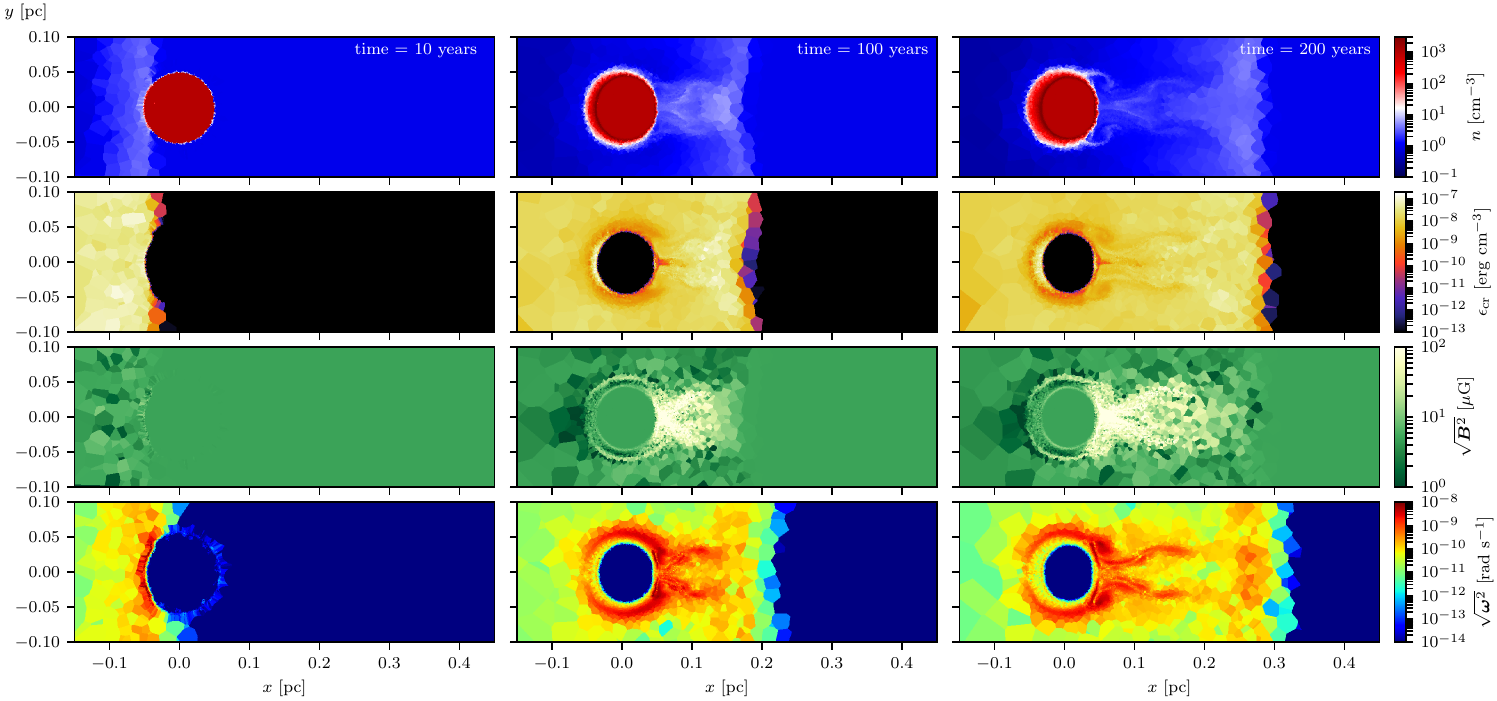}
  \end{center}
  \caption{Snapshots of the shock-cloud interaction at $t=10, 100$ and
      $200$~yrs after the collision for a density contrast of $\delta = 10^3$. For each
      time we show slices of the gas density (top row), CR energy density (second row),
      magnetic field strength (third row) and magnitude of the vorticity (fourth
      row).} 
  \label{fig: shocktube}
\end{figure*}

We use the code \AREPO, which employs an unstructured mesh that is defined as
  the Voronoi tessellation of a set of mesh-generating points. If they move with
  the local fluid speed, the scheme inherits the advantages of Lagrangian fluid
  methods that keep the mass per cell approximately fixed. \AREPO allows for
  super-Lagrangian resolution capabilities by inserting new mesh points.  Hence,
  in order to accurately describe the dynamics of the dense clouds, each
  small clump is resolved with $10^3$ cells that are uniformly distributed
  within a spherical volume. To ensure a smooth transition between the larger
  cell size in the low-density diffuse ISM and the well-resolved clumps, we
  initially place a buffer shell of thickness $0.01~\pc$ with the density of the
  diffuse ISM and the high resolution of the clump cells. Despite the high
  post-shock vorticity caused by the impact of shock front on the clumps, the
  volume fraction occupied by the dense bullets is small enough to maintain the
  shock front practically self similar throughout its evolution in the medium.

\subsection{Interaction of a shock with a dense clump}

Because of momentum conservation, the SNR shock penetrates much slower in the
  dense clump in comparison to the shock velocity in the dilute medium outside
  the clump and thus, only a fraction of dense gas inside these objects is
  accelerated at a given time.  To investigate the mass fraction $\eta$ that is
  processed by the shock, we conduct simulations with clumps of different
  density ($n_\rmn{c} = \{10^2, 10^3, 10^4\}~\cm^{-3}$). To accurately describe
  the dynamics, we resolve the cloud with $1.5 \times 10^5$ particles within a
  radius of $0.05~\pc$. We place the cloud in a shock-tube of length $2~\pc$ and
  density of $n_0 = 1~\cm^{-3}$. The shock is set to propagate in the $x$
  direction with initial density and pressure jumps so that the initial shock
  velocity is $v_\mathrm{s} = 5000~\km~\s^{-1}$.  Different snapshots of the
  shock-cloud interaction for a density contrast of $\delta = 10^3$ are reported
  in Fig.~\ref{fig: shocktube}.

When the supernova blast wave impacts a dense structure at some constant
  (oblique) angle, it experiences the same ``shock deflection'' and amount of
  entropy injection along the shock front because only the parallel velocity
  component is reduced by the compression ratio while the perpendicular
  component is conserved. If the blast wave encounters a dense, curved
  structure, the incoming velocity experiences a different amount of deflection
  along the shock surface (perpendicular to the shock normal), which injects
  voriticty according to \citeauthor{1937ZaMM...17....1C}’s theorem
  (\citeyear{1937ZaMM...17....1C}). This vorticity is injected at the length
  scale of the clump and cascades to smaller scales with increasing distance
  from the shock. The shock in the dilute phase closes in after passing the
  clump and eventually straightens up as it closes the dip at the location of
  the clump after about 100 years (see Fig.~\ref{fig: shocktube}). While the
  shock front propagates seemingly undisturbed at larger distances from the
  clump, it has dramatically slowed down inside the clump as a result of
  momentum conservation. The relic of such a complex history of the shock
  evolution is a highly turbulent tail in the downstream of the clump, which is
  able to drive a small-scale turbulent dynamo that amplifies the magnetic field
  there. This behavior is consistent with the lower resolution simulation of the
  entire SNR presented in top right panel of Fig.~\ref{fig:comparison1}, where a
  system of uniformly distributed clumps leave several magnetized finger-like
  structures after the collision with the blast wave. 

\begin{figure}
  \begin{center}
    \includegraphics[scale=0.58]{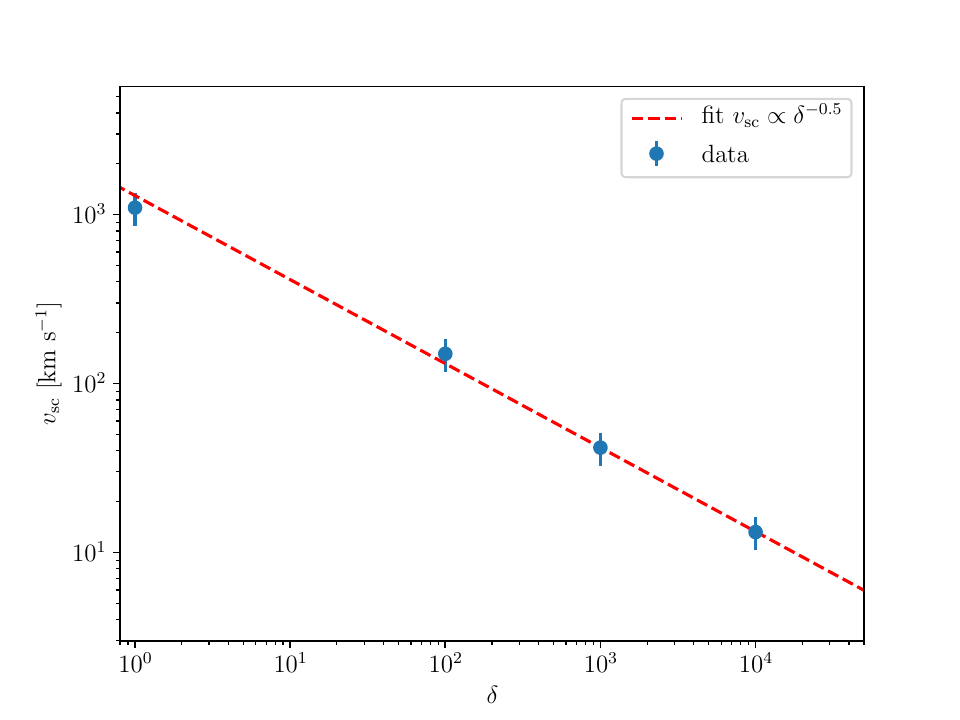}
  \end{center}
  \caption{Simulated shock velocity $v_\rmn{sc}$ inside a dense clump as a
      function of the density contrast for the cases $\delta =  [10^2,
        10^3, 10^4]$. The point for $\delta =1$ represents the shock velocity in
      the dilute phase. Within the uncertainty, the simulation points follow the
      $\delta^{-1/2}$ scaling behaviour derived in Eq.~\eqref{eq: mu2}. }
  \label{fig: delta}
\end{figure}

\begin{figure}
  \begin{center}
    \includegraphics[scale=0.55]{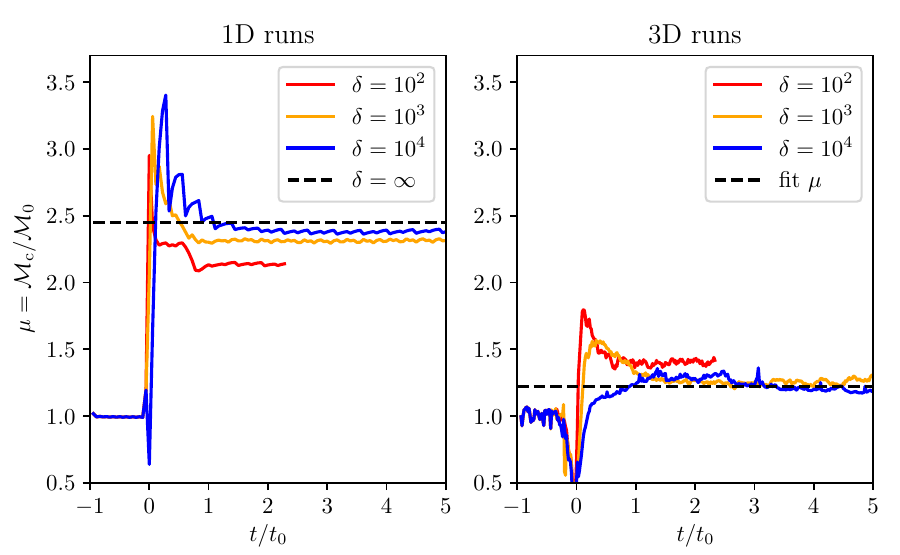}
  \end{center}
  \caption{Mach number ratio evolution after the interaction with a dense region for various density contrasts. \emph{Left}: the 1D shocktube runs show a good agreement with the theoretical value reported in Eq.~\eqref{eq: mu2} (dashed line).  \emph{Right}: the 3D runs show a lower pre-factor with respect to the 1D theoretical estimate. The case with $\delta = 10^2$ is interrupted in both case by the finite size of the clumpy region.}
  \label{fig: mach}
\end{figure}

The interaction of a shock with a spherical object has been previously
  studied by \citet{2011ApJ...730...22P} for a low-density bubble, the results
  of which can also be applied to the case of a dense clump by exchanging the
  rarefaction wave with a reverse shock characteristics that propagates in the
  opposite direction of the original shock. Solving the Riemann problem in this
  case, we can relate the initial shock speed $v_{\rmn{s0}}$ in the dilute phase
  to the shock speed $v_{\mathrm{sc}}$ propagating inside the clump. The dilute
  and the clump phases have the same initial pressure $P_0$ while the sound
  speeds of the dense and dilute media are related by $c_\mathrm{c} =
  \sqrt{\gamma P_0 / \rho_\mathrm{c} }= c_0 / \sqrt{\delta}$.  The equation for
  a left reverse shock condition \citep{Toro} reads: 
\begin{equation}
\label{eq: starL}
v_\ast = v_\ls + (P_\ls - P_\ast) \sqrt{\dfrac{2}{\gamma_\ls+1}}\left[ \rho_\mathrm{L} \left(P_\ast + \dfrac{\gamma_\ls-1}{\gamma_\ls+1} P_\mathrm{L}\right) \right]^{-1/2}
\end{equation}
where the subscript ``$\ast$'' denotes the state behind the reverse shock
  while the subscript L represents the state ahead of the reverse shock.
Introducing the Mach number ratio $\mu = \mathcal{M}_\mathrm{c} /
  \mathcal{M}_0 = \sqrt{\delta}\, v_\mathrm{sc} / v_{\rmn{s0}}$, assuming
  $\gamma_0=\gamma_\rmn{c}=5/3$, and combining the various jump conditions (see
  eq. (B1) in Appendix B of \citealt{2011ApJ...730...22P} with subscript
  $\rmn{L} = 3$) we find an equation for $\mu$ as a function of the Mach number
  $\mathcal{M}_0$ in the dilute phase and the density contrast $\delta$:
\begin{equation}
\label{eq: mu1}
\begin{aligned}
1 = &\sqrt{\delta } \mu - \mu  \mathcal{M}_0^2 \left(\sqrt{\delta }-\mu \right) \\ 
&+ \mu  \left(\mu^2-1\right) \mathcal{M}_0^2 \sqrt{\frac{\delta  \left(\mathcal{M}_0^2+3\right)}{ \left(4 \mu^2+1\right) \mathcal{M}_0^2-1}}, 
\end{aligned}
\end{equation}
which has a numerical solution for $\mu$ given $\mathcal{M}_0$ and
  $\delta$. In the regime of strong shocks (i.e., $\mathcal{M}_0 \gg 1$)
  Eq.~\eqref{eq: mu1} is reduced to the simpler form:
\begin{equation}
\label{eq: mu2}
  \mu = (1 - \mu^2) \sqrt{\frac{\delta }{4 \mu^2+1}}+\sqrt{\delta }
  \end{equation}  
For $\delta \gg 1$ Eq.~\eqref{eq: mu2} can be easily solved and has the
  positive root $\mu=\sqrt{6}$. In order to account for multi-dimensional
  effects of order unity, we introduce a factor $f_\mathrm{d}$ that we calibrate
  on our three-dimensional (3D) simulations and write the solutions as $\mu =
  f_\mathrm{d} \sqrt{6}$. Because $\mu = \sqrt{\delta}\, v_\mathrm{sc} /
  v_{\rmn{s0}}$ we conclude that $v_{\mathrm{sc}} \propto \delta^{-1/2}
  v_{\rmn{s0}} $. We confirm the applicability of this scaling behavior in 3D
  simulations by fitting the appropriate power law to the shock tube simulation
  of the three different density contrasts (see Fig.~\ref{fig: delta}). The
  theoretically expected scaling matches the simulations within the variance of
  the shock velocity, however the pre-factor found in our 3D simulations is
  closer to 1.2 rather than 2, which means that $f_\mathrm{d} \simeq 1/2$. A
  comparison between 1D runs and 3D runs for density contrasts of $10^2$, $10^3$
  and $10^4$ is shown in Fig.~\ref{fig: mach}. In the following formulas to
  construct the analytical model for the gamma-ray luminosity we will apply the
  3D pre-factor rather than the 1D prediction.

\subsection{Constructing the analytical model}

To calculate the mass fraction of the clump that is processed by the shock,
  we proceed as follows.  We follow the shock-wave propagation inside the clump
  and calculate the amount of shocked gas inside the clump as a function of
  time. We define the shock-processed clump mass fraction $\eta(t)$ as the ratio
  between the shocked gas mass of the dense clump and its total mass as a
  function of time. 

A shock propagates in a shock-tube with the velocity $\propto t^{-1/3}$ as
  predicted by the 1D Sedov problem while a 3D blast wave evolves according to
  $t^{-3/5}$. In the latter case we integrate the velocity to obtain the shock
  propagation length inside the clump as a function of time:
\begin{equation}
\label{eq: radius_clump}
\begin{aligned}
r(t,t_0) &= f_\mathrm{d} \sqrt{\frac{6}{\delta}} \int_{t_0}^{t} v_{\mathrm{s0}}(t') \de t'  \\
 & =f_\mathrm{d}  \sqrt{\frac{6}{\delta}} \left( \dfrac{E_{\mathrm{SN}}}{\alpha\rho_0} \right)^{1/5} [ t^{2/5} - t_0^{2/5}]
\end{aligned}
\end{equation}
which is valid for $t_0 < t < t_\mathrm{c}(t_0)$, where $t_0$ is the time of
  the supernova shock impacting the clump and $t_\mathrm{c}$ is the crossing
  time of the shock inside the clump, which can be obtained from the requirement
  that the shock propagation length has to be smaller than the size of the
  clump,
\begin{equation}
t_\mathrm{c}(t_0) = \left[ \dfrac{2 R_\mathrm{c}}{f_\mathrm{d} \sqrt{6}} \delta^{1/2}  \left( \dfrac{E_{\mathrm{SN}}}{\alpha\rho_0} \right)^{-1/5} + t_0^{2/5} \right]^{5/2}.
\end{equation}
This means that for typical values, the shock propagation length in the
  clump is
\begin{equation}
\label{eq: radius_clump_2}
r(t,t_0) = 0.064~\pc~\left(\frac{\delta}{10^3}\right)^{-1/2}  \left( \dfrac{n }{0.1~\cm^{-3}}\right)^{-1/5}
\end{equation}
for $t=1~\kyr$ and $t_0=500~\yr$. We can determine the volume of the shocked
  material inside the clump using Eq.~\eqref{eq: radius_clump} as the height of
  a polar cap as a function of time, while neglecting the curvature of the shock
  inside the clump. The ratio between the shocked polar cap volume of the clump
  and its total volume reads
\begin{equation}
\label{eq: efficiency_single}
\eta(t,t_0) = \dfrac{V_{\mathrm{cap}}}{V_{\mathrm{tot}}}
= \dfrac{\pi/3\times r^2 (3 R_\mathrm{c} - r)}{4\pi/3 \times R^3_\mathrm{c} }
= \dfrac{r^2(3 R_\mathrm{c} - r)}{4 R^3_\mathrm{c}},
\end{equation}
where $r=r(t, t_0)$ and we used the fact that the clump density is
  constant. Eq.~\eqref{eq: efficiency_single} represents the mass fraction
  accelerated by the shock as a function of time. For $r(t,t_0) =
  2R_\mathrm{c}$ we set $\eta(t,t_0) = 0$ to indicate that the clump 
  is either destroyed by MHD instabilities or emptied of CRs.

Assuming that a uniform spatial distribution of clumps is overrun by a
  self-similar quasi-spherical blast wave, it is clear that the shock interacts
  at different times with individual clumps. This implies that the innermost
  clumps close to the explosion site have a higher fraction of shocked mass in
  comparison to clumps at the periphery. For this reason we have to weight
  Eq.~\eqref{eq: efficiency_single} with a distribution of clumps that interact
  with the shock per unit time. Assuming that the number of clumps is large so
  that the continuum limit applies, the number of clumps that start to interact
  with the SNR shock is given by
\begin{equation}
\label{eq: clump_num}
\begin{aligned}
\dot{N}(t) &= \dfrac{\de N(t)}{\de t}  =  \dfrac{\langle \rho_c \rangle}{M_\mathrm{c}}  \dfrac{\de V_{s}(t)}{\de t} = \dfrac{4}{3} \pi   \dfrac{\langle \rho_c \rangle}{M_\mathrm{c}}  \dfrac{\de [r_s^3 (t)]}{\de t}  \\
& =\dfrac{1.7}{\yr}\left( \dfrac{t}{1~\kyr}\right)^{1/5} \left(\dfrac{\langle \rho_c \rangle}{M_\odot \pc^{-3}} \right) \left( \dfrac{M_{\mathrm{c}}}{M_\odot}\right)^{-1}
\end{aligned}
\end{equation}
In the last step, we assume the Sedov-Taylor regime for the SNR because the
  fraction of volume swept-up in the free expansion phase is only about $0.3\%$
  of the final volume and it encompasses only less than $20$ clumps for our
  setup. Using Eqs.~\eqref{eq: efficiency_single} and \eqref{eq: clump_num} the
  average efficiency $\eta(t)$ weighted by the number of interacting clumps per
  unit time for a SNR reads
\begin{equation}
\label{eq: eta_t}
\begin{aligned}
\bar{\eta}(t) &= \dfrac{\displaystyle\int_0^t \dot{N}(t') \eta(t,t') \de t'}{\displaystyle\int_0^t \dot{N}(t')\de t'} \\   
& \simeq \left\lbrace  \dfrac{1}{52 \tilde{r}^{9/4}(t,\delta,n)} +   7 \tilde{r}^{2/3}(t,\delta,n) \right\rbrace^{-1} ,
\end{aligned}
\end{equation}
where
\begin{equation}
\tilde{r}(t,\delta,n) = 0.25 \left(\dfrac{\delta}{10^3}\right)^{-1/2} \left( \dfrac{n}{0.1~\cm^{-3}} \right)^{-1/5} \left( \dfrac{t}{1~\kyr} \right)^{2/5}.
\end{equation}
Results for different values of $\delta$ are plotted in Fig.~\ref{fig:
    etat}. For $\delta = 3\times 10^3$ and $t = 2.5~\kyr$ we find $\eta(t)
  \simeq 32\%$, which is slightly higher than the value found in the simulation
  (around $30\%$). 
  
\begin{figure}
  \begin{center}
    \includegraphics[scale=0.58]{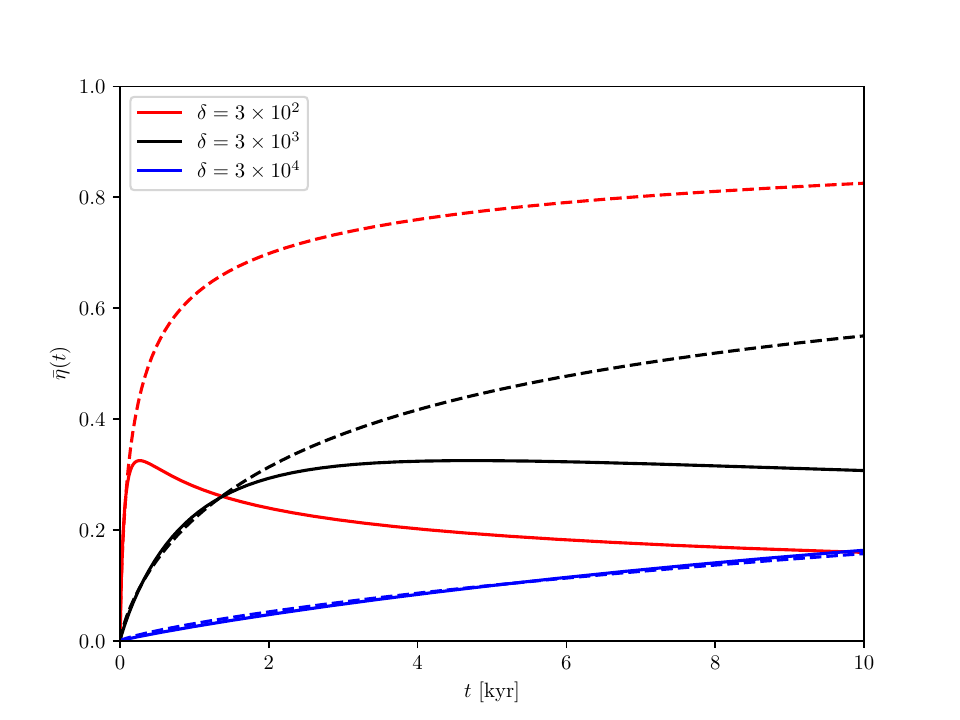}
  \end{center}
  \caption{Evolution of $\bar{\eta}(t)$ for different values of $\delta = 3 \times [10^2, 10^3, 10^4]$ for $n = 0.42~\cm^{-3}$. The lines represent the theoretical values calculated via Eq.~\eqref{eq: eta_t}. Although an increasing number of clumps are hit by the shock the slowdown of the blast wave transfers a decreasing amount of momentum to the clumps leading to an ideal asymptotic value for $\bar{\eta}(t)$. The dashed lines represent the case where the clumps are not destroyed and continue to be active sources of CRs.}
  \label{fig: etat}
\end{figure}

To obtain the gamma-ray luminosity $\mathcal{L}_\gamma = 4 \pi D^2
  \mathcal{F}_\gamma$ due to hadronic CR proton interactions, we separately
  calculate the contributions from the dilute phase and the clouds and then add
  them together:
\begin{equation}
  \label{eq: luminosity sum}
\mathcal{L}_{\gamma} =\int s_{\pi^0 \rightarrow \gamma} \de V=
\mathcal{L}_{\gamma, \mathrm{d}} + \mathcal{L}_{\gamma, \mathrm{c}},
\end{equation}
where $s_{\pi^0 \rightarrow \gamma}$ is the neutral pion-decay source function
  \citep{2004A&A...413...17P}. The contribution from the dilute phase is given by
  \begin{equation}
\label{eq: luminosity}
\begin{aligned}
  \mathcal{L}_{\gamma, \mathrm{d}} = 5.3 \times 10^{32} \left( \dfrac{W_\mathrm{p}}{10^{50}~\erg} \right) \left( \dfrac{\langle n\rangle}{0.1~\cm^{-3}} \right) \dfrac{\ph}{\s},
\end{aligned}
\end{equation}
where $W_\mathrm{p}$ the total proton energy integrated over the whole SNR
    volume and $\langle n\rangle$ is the volume averaged number density, which
    approximately coincides with the number density on the dilute phase due to
    the negligible volume filling factor of the dense clumps, $f_V =
    V_{\mathrm{c}} / V= 2.4 \times 10^{-4}$ for our
    parameters. The gamma-ray luminosity associated to the clumped gas is:
\begin{equation}
\label{eq: Lgamma_clump}
\begin{aligned}
  &\mathcal{L}_{\gamma, \mathrm{c}}  = \int s_{\pi^0 \rightarrow \gamma} \de V_{\mathrm{c}} =  \int s_{\pi^0 \rightarrow \gamma}  f_V \de V  \\
  & = 1.6 \times 10^{32} \left(\dfrac{\bar{\eta}(t)}{30\%} \right) \left( \dfrac{f_V}{10^{-4}}\right) \left( \dfrac{n_\mathrm{c}}{10^3~\cm^{-3}} \right) \left( \dfrac{W_\mathrm{p}}{10^{50}~\erg} \right)  \dfrac{\ph}{\s}
\end{aligned}
\end{equation}
where $V_\mathrm{c}$ is the volume occupied by the clumps and $\bar{\eta}(t)$
  is the average fraction of the shocked clump volume fraction inside the
  remnant volume. In our setup we set $n_\mathrm{c} = 1.4 \times 10^3
  \cm^{-3}$. Combining  Eqs.~\eqref{eq: luminosity} and \eqref{eq: Lgamma_clump}
  we get
\begin{equation}
\begin{aligned}
\mathcal{L}_{\gamma} &= \mathcal{L}_{\gamma, \mathrm{s}}  + \mathcal{L}_{\gamma, \mathrm{c}}  
\\   & = 5.3 \times 10^{32} (1+ \chi)   \left( \dfrac{W_\mathrm{p}}{10^{50}~\erg} \right) \left( \dfrac{n}{0.1~\cm^{-3}} \right) \dfrac{\ph}{\s}
\end{aligned}
\end{equation}
where
\begin{equation}
\label{eq: chi_l}
\chi = 0.3 \left(\dfrac{\bar{\eta}(t)}{30\%} \right) \left( \dfrac{f_V}{10^{-4}}\right) \left( \dfrac{n_\mathrm{c}}{10^3~\cm^{-3}} \right) \left( \dfrac{n}{0.1~\cm^{-3}} \right)^{-1}.
\end{equation}
It is easy to verify that Eq.~\eqref{eq: chi factor} is equivalent to
  Eq.~\eqref{eq: chi_l} because of the following identity: 
 \begin{equation}
 f_V n_\mathrm{c} =  \dfrac{ f_V \rho_\mathrm{c}  }{\mu_{\mathrm{H}} m_\mathrm{p}}   =\dfrac{ \langle \rho_\mathrm{c} \rangle }{\mu_{\mathrm{H}} m_\mathrm{p}}  
 \end{equation}
 In terms of normalized quantities the previous equation becomes:
\begin{equation}
 \left( \dfrac{f_V}{2.4 \times 10^{-4}}\right) \left( \dfrac{n_\mathrm{c}}{1.4 \times 10^3~\cm^{-3}} \right) = \left( \dfrac{\langle \rho_\mathrm{c}\rangle}{1.6 \times 10^{-2} M_\odot \pc^{-3}} \right) .
\end{equation}

\end{document}